\documentclass[useAMS,usenatbib]{mn2e}
\usepackage{graphicx}
\usepackage{natbib,float,longtable,subfigure}
\usepackage{amsmath,amssymb,lscape}
\usepackage{mathptmx}
\usepackage{caption}
\usepackage{color}
\newcommand{\todo}{\ifmmode \text{\Huge{\(\bullet\)}} \else {\Huge$\bullet$}\fi}
\newcommand{\tido}{\ifmmode {\bullet} \else $\bullet$\fi}

\newcommand{\E        }[1]{\ifmmode 10^{#1} \else $10^{#1}$\fi}
\newcommand{\tE        }[1]{\ifmmode \times10^{#1} \else $\times10^{#1}$\fi}
\newcommand{\til}{\ifmmode \sim \else $\sim$\fi}
\renewcommand{\~} {\ifmmode \sim \else $\sim$\fi}

\newcommand{\pc}	{\ifmmode {\rm pc} \else pc\fi}
\newcommand{\ld}	{\ifmmode {\rm l.d.} \else l.d.\fi}
\newcommand{\kms}	{\ifmmode {\rm km\,s}^{-1} \else km\,s$^{-1}$\fi}
\newcommand{\cc}	{\ifmmode {\rm cm}^{-3}    \else cm$^{-3}$\fi}
\newcommand{\cmii}	{\ifmmode {\rm cm}^{-2}    \else cm$^{-2}$\fi}
\newcommand{\ergs}	{\ifmmode {\rm erg\,s}^{-1} \else erg s$^{-1}$\fi}
\newcommand{\ergcms}	{\ifmmode {\rm erg\,cm}^{-2}\,{\rm s}^{-1} \else erg\,cm$^{-2}$\,s$^{-1}$\fi}
\newcommand{\ergcmsA}	{\ifmmode {\rm erg\,cm}^{-2}\,{\rm s}^{-1}\,{\rm\AA}^{-1}
\else erg\,cm$^{-2}$\,s$^{-1}$\,\AA$^{-1}$\fi}
\newcommand{  \ergcmsHz  }{\ifmmode{\rm erg\,cm}^{-2}\,{\rm s}^{-1}\,{\rm Hz}^{-1}
                       \else ergs\,cm$^{-2}$\,s$^{-1}$\,Hz$^{-1}$\fi}
\newcommand{\kev}	{\ifmmode {\rm keV} \else keV\fi}

\newcommand{\mic}	{\ifmmode {\rm \mu m} \else $\mu$m\fi}
\newcommand{\vFWHM}	{\ifmmode v_{\mbox{\tiny FWHM}} \else $v_{\mbox{\tiny FWHM}}$\fi}
\newcommand{\vBLR}	{\ifmmode v_{\mbox{\tiny BLR}} \else $v_{\mbox{\tiny BLR}}$\fi}
\newcommand{\sigBLR}	{\ifmmode \sigma_{\mbox{\tiny BLR}} \else $\sigma_{\mbox{\tiny BLR}}$\fi}
\newcommand{\vNLR}	{\ifmmode v_{\mbox{\tiny NLR}} \else $v_{\mbox{\tiny NLR}}$\fi}
\newcommand{\tauBLR}	{\ifmmode \tau_{\mbox{\tiny BLR}} \else $\tau_{\mbox{\tiny BLR}}$\fi}

\newcommand{\Hubble}	{\ifmmode {\rm km\,s}^{-1}\,{\rm Mpc}^{-1} \else km\,s$^{-1}$\,Mpc$^{-1}$\fi}
\newcommand{\NDunit}	{\ifmmode {\rm Mpc}^{-3} \else Mpc$^{-3}$\fi}
\newcommand{\LFunit}	{\ifmmode {\rm Mpc}^{-3}\,{\rm mag}^{-1} \else Mpc$^{-3}$\,mag$^{-1}$\fi}
\newcommand{\MFunit}	{\ifmmode {\rm Mpc}^{-3}\,{\rm dex}^{-1} \else Mpc$^{-3}$\,dex$^{-1}$\fi}

\newcommand{\Msun}{\ifmmode M_{\odot} \else $M_{\odot}$\fi}
\newcommand{\Lsun}{\ifmmode L_{\odot} \else $L_{\odot}$\fi}
\newcommand{\Zsun}{\ifmmode Z_{\odot} \else $Z_{\odot}$\fi}
\newcommand{\mpyr}{\ifmmode \Msun\,{\rm yr}^{-1} \else $\Msun\,{\rm yr}^{-1}$\fi}

\newcommand{\qnote}{\ifmmode q_{0} \else $q_{0}$\fi}
\newcommand{\Hnote}{\ifmmode H_{0} \else $H_{0}$\fi}
\newcommand{\hnote}{\ifmmode h_{0} \else $h_{0}$\fi}
\newcommand{\anote}{\ifmmode a_{0} \else $a_{0}$\fi}
\newcommand{\tnote}{\ifmmode t_{0} \else $t_{0}$\fi}

\newcommand{  \Halpha   }{\ifmmode {\rm H}\alpha \else H$\alpha$\fi}
\newcommand{  \ha   	}{\ifmmode {\rm H}\alpha \else H$\alpha$\fi}
\newcommand{  \Hbeta    }{\ifmmode {\rm H}\beta \else H$\beta$\fi}
\newcommand{  \hb    	}{\ifmmode {\rm H}\beta \else H$\beta$\fi}
\newcommand{  \Hgamma   }{\ifmmode {\rm H}\gamma \else H$\gamma$\fi}
\newcommand{  \Hdelta   }{\ifmmode {\rm H}\delta \else H$\delta$\fi}
\newcommand{  \Lya      }{\ifmmode {\rm Ly}\alpha \else Ly$\alpha$\fi}
\newcommand{  \Lyb      }{\ifmmode {\rm Ly}\beta \else Ly$\beta$\fi}
\newcommand{  \Pa       }{\ifmmode {\rm P}\alpha \else P$\alpha$\fi}
\newcommand{  \Pb       }{\ifmmode {\rm P}\beta \else P$\beta$\fi}
\newcommand{  \Bra      }{\ifmmode {\rm Br}\alpha \else Br$\alpha$\fi}
\newcommand{  \Brg      }{\ifmmode {\rm Br}\gamma \else Br$\gamma$\fi}
\newcommand{  \hii      }{\ifmmode {\rm H}\,\textsc{ii} \else H\,\textsc{ii}\fi}
\newcommand{  \hei      }{\ifmmode {\rm He}\,\textsc{i} \else He\,\textsc{i}\fi}
\newcommand{  \heii     }{\ifmmode {\rm He}\,\textsc{ii} \else He\,\textsc{ii}\fi}
\newcommand{  \HeIIuv   }{\ifmmode {\rm He}\,\textsc{ii}\,\lambda1640 \else He\,\textsc{ii}\,$\lambda1640$\fi}
\newcommand{  \HeIIop   }{\ifmmode {\rm He}\,\textsc{ii}\,\lambda4686 \else He\,\textsc{ii}\,$\lambda4686$\fi}
\newcommand{  \cii      }{\ifmmode {\rm C}\,\textsc{ii}  \else C\,\textsc{ii}\fi}
\newcommand{  \ciii     }{\ifmmode {\rm C}\,\textsc{iii}\right] \else C\,\textsc{iii}]\fi}
\newcommand{  \CIII     }{\ifmmode {\rm C}\,\textsc{iii}\right]\,\lambda1909 \else C\,\textsc{iii}]\,$\lambda1909$\fi}
\newcommand{  \civ      }{\ifmmode {\rm C}\,\textsc{iv}  \else C\,\textsc{iv}\fi}
\newcommand{  \CIV      }{\ifmmode {\rm C}\,\textsc{iv}\,\lambda1549 \else C\,\textsc{iv}\,$\lambda1549$\fi}
\newcommand{  \niii     }{\ifmmode {\rm N}\,\textsc{iii} \else N\,\textsc{iii}\fi}
\newcommand{  \niv      }{\ifmmode {\rm N}\,\textsc{iv}  \else N\,\textsc{iv}\fi}
\newcommand{  \NIVuv    }{\ifmmode {\rm N}\,\textsc{iv}\,\lambda1486 \else N\,\textsc{iv}\,$\lambda1486$\fi}
\newcommand{  \nv       }{\ifmmode {\rm N}\,\textsc{v}   \else N\,\textsc{v}\fi}
\newcommand{\oi}{\ifmmode \left[{\rm O}\,\textsc{i}\right] \else [O\,{\sc i}]\fi}
\newcommand{\OI}{\ifmmode \left[{\rm O}\,\textsc{i}\right]\,\lambda6300 \else [O\,{\sc i}]$\,\lambda6300$\fi}

\newcommand{\oii}{\ifmmode \left[{\rm O}\,\textsc{ii}\right] \else [O\,{\sc ii}]\fi}
\newcommand{\OII}{\ifmmode \left[{\rm O}\,\textsc{ii}\right]\,\lambda3727 \else [O\,{\sc ii}]\,$\lambda3727$\fi}
\newcommand{\oiii}{\ifmmode \left[{\rm O}\,\textsc{iii}\right] \else [O\,{\sc iii}]\fi}
\newcommand{\OIII}{\ifmmode \left[{\rm O}\,\textsc{iii}\right]\,\lambda5007 \else [O\,{\sc iii}]\,$\lambda5007$\fi}

\newcommand{\NII}{\ifmmode \left[{\rm N}\,\textsc{ii}\right]\,\lambda6583 \else [N\,{\sc ii}]$\,\lambda6583$\fi}

\newcommand{\NeIII}{\ifmmode \left[{\rm Ne}\,\textsc{iii}\right]\,\lambda3968 \else [Ne\,{\sc iii}]$\,\lambda3968$\fi}

\newcommand{\NeV}{\ifmmode \left[{\rm Ne}\,\textsc{v}\right]\,\lambda3426 \else [Ne\,{\sc v}]$\,\lambda3426$\fi}

\newcommand{\HeII}{\ifmmode \left[{\rm He}\,\textsc{ii}\right]\,\lambda4686 \else [He\,{\sc v}]$\,\lambda4686$\fi}

\newcommand{\sii}{\ifmmode \left[{\rm S}\,\textsc{ii}\right] \else [S\,{\sc ii}]\fi}

\newcommand{\SII}{\ifmmode \left[{\rm S}\,\textsc{ii}\right]\,\lambda6717,6731 \else [S\,{\sc ii}]$\,\lambda6717,6731$\fi}

\newcommand{\nii}{\ifmmode \left[{\rm N}\,\textsc{ii}\right] \else [N\,{\sc ii}]\fi}

\newcommand{  \OIIIuv   }{\ifmmode {\rm O}\,\textsc{iii}\,\lambda1663 \else O\,\textsc{iii}\,$\lambda1663$\fi}
\newcommand{  \oiv      }{\ifmmode {\rm O}\,\textsc{iv}  \else O\,\textsc{iv}\fi}
\newcommand{  \OIVuv    }{\ifmmode {\rm O}\,\textsc{iv}\,\lambda1402  \else O\,\textsc{iv}\,$\lambda1402$\fi}
\newcommand{  \OIVIR    }{\ifmmode {\rm O}\,\textsc{iv}\,25.9\,\mu {\rm m} \else O\,\textsc{iv}\,$25.9\,\mu$m\fi}
\newcommand{  \ovi      }{\ifmmode {\rm O}\,\textsc{vi}   \else O\,\textsc{vi}\fi}
\newcommand{  \Ovi      }{\ifmmode {\rm O}\,\textsc{vi}\,\lambda1035 \else O\,\textsc{vi}\,$\lambda1035$\fi}
\newcommand{  \nei      }{\ifmmode {\rm Ne}\,\textsc{i}   \else Ne\,\textsc{i}\fi}
\newcommand{  \neii     }{\ifmmode {\rm Ne}\,\textsc{ii}  \else Ne\,\textsc{ii}\fi}
\newcommand{  \NeiiIR   }{\ifmmode {\rm Ne}\,\textsc{ii}\,12.8\,\mu {\rm m} \else Ne\,\textsc{ii}\,$12.8\,\mu$m\fi}
\newcommand{  \neiii    }{\ifmmode {\rm Ne}\,\textsc{iii} \else [Ne\,\textsc{iii}]\fi}
\newcommand{  \neiv     }{\ifmmode {\rm Ne}\,\textsc{iv}  \else Ne\,\textsc{iv}\fi}
\newcommand{  \nev      }{\ifmmode {\rm Ne}\,\textsc{v}   \else [Ne\,\textsc{v}]\fi}
\newcommand{  \NevIR    }{\ifmmode {\rm Ne}\,\textsc{v}\,24.3\,\mu {\rm m} \else Ne\,\textsc{v}\,$24.3\,\mu$m\fi}
\newcommand{  \nevi     }{\ifmmode {\rm Ne}\,\textsc{vi}  \else Ne\,\textsc{vi}\fi}
\newcommand{  \mgi      }{\ifmmode {\rm Mg}\,\textsc{i}   \else Mg\,\textsc{i}\fi}
\newcommand{  \mgii     }{\ifmmode {\rm Mg}\,\textsc{ii}  \else Mg\,\textsc{ii}\fi}
\newcommand{  \MgII     }{\ifmmode {\rm Mg}\,\textsc{ii}\,\lambda2798 \else Mg\,\textsc{ii}\,$\lambda2798$\fi}

\newcommand{  \siii     }{\ifmmode {\rm S}\,\textsc{iii} \else S\,\textsc{iii}\fi}
\newcommand{  \siv      }{\ifmmode {\rm S}\,\textsc{iv}  \else S\,\textsc{iv}\fi}
\newcommand{  \sili     }{\ifmmode {\rm Si}\,\textsc{i}   \else Si\,\textsc{i}\fi}
\newcommand{  \silii    }{\ifmmode {\rm Si}\,\textsc{ii}  \else Si\,\textsc{ii}\fi}
\newcommand{  \Siliv    }{\ifmmode {\rm Si}\,\textsc{iv}  \else Si\,\textsc{iv}\fi}
\newcommand{  \SilIVuv  }{\ifmmode {\rm Si}\,\textsc{iv}\,\lambda1400  \else Si\,\textsc{iv}\,$\lambda1400$\fi}

\newcommand{  \caii     }{\ifmmode {\rm Ca}\,\textsc{ii}   \else Ca\,\textsc{ii}\fi}
\newcommand{  \feii     }{\ifmmode {\rm Fe}\,\textsc{ii}  \else Fe\,\textsc{ii}\fi}
\newcommand{  \feiii    }{\ifmmode {\rm Fe}\,\textsc{iii} \else Fe\,\textsc{iii}\fi}

\newcommand{ \Lhb   }{\ifmmode L\left(\hb\right) \else $L\left(\hb\right)$\fi}
\newcommand{ \fwhb  }{\ifmmode {\rm FWHM}\left(\hb\right) \else FWHM(\hb)\fi}
\newcommand{ \Lmg   }{\ifmmode L\left(\mgii\right) \else $L\left(\mgii\right)$\fi}
\newcommand{ \fwmg  }{\ifmmode {\rm FWHM}\left(\mgii\right) \else FWHM(\mgii)\fi}
\newcommand{ \Lciv  }{\ifmmode L\left(\civ\right) \else $L\left(\civ\right)$\fi}
\newcommand{ \fwciv }{\ifmmode {\rm FWHM}\left(\civ\right) \else FWHM(\civ)\fi}
\newcommand{ \fwhm  }{\ifmmode {\rm FWHM} \else FWHM\fi} 
\newcommand{ \voff  }{\ifmmode v_{\rm off} \else $v_{\rm off}$\fi} 

\newcommand{ \mumg  }{\ifmmode \mu\left(\mgii\right) \else $\mu\left(\mgii\right)$\fi}
\newcommand{ \fmg   }{\ifmmode f\left(\mgii\right) \else $f\left(\mgii\right)$\fi}
\newcommand{ \muciv }{\ifmmode \mu\left(\civ\right) \else $\mu\left(\civ\right)$\fi}
\newcommand{ \fciv  }{\ifmmode f\left(\civ\right) \else $f\left(\civ\right)$\fi}

\newcommand{  \auvo     }{\ifmmode \alpha_{\nu,{\rm UVO}} \else $\alpha_{\nu,{\rm UVO}}$\fi}
\newcommand{  \Ledd     }{\ifmmode L_{\rm Edd} \else $L_{\rm Edd}$\fi}
\newcommand{  \lamLlam  }{\ifmmode \lambda L_{\lambda} \else $\lambda L_{\lambda}$\fi}
\newcommand{  \lLl      }{\ifmmode \lambda L_{\lambda} \else $\lambda L_{\lambda}$\fi}
\newcommand{  \nuLnu    }{\ifmmode \nu L_{\nu} \else $\nu L_{\nu}$\fi}
\newcommand{  \nLn      }{\ifmmode \nu L_{\nu} \else $\nu L_{\nu}$\fi}
\newcommand{  \Luv      }{\ifmmode L_{1450} \else $L_{1450}$\fi}
\newcommand{  \Lop      }{\ifmmode L_{5100} \else $L_{5100}$\fi}
\newcommand{  \lLop     }{\ifmmode \log\left(\Lop/\ergs\right) \else $\log\left(\Lop/\ergs\right)$\fi}
\newcommand{  \Lthree   }{\ifmmode L_{3000} \else $L_{3000}$\fi}
\newcommand{  \lLthree  }{\ifmmode \log\left(\Lthree/\ergs\right) \else $\log\left(\Lthree/\ergs\right)$\fi}

\newcommand{\Fthree}{\ifmmode F_{3000} \else $F_{3000}$\fi}
\newcommand{\fuv}{\ifmmode f_{\lambda}\left(1450{\rm \AA}\right) \else $f_{\lambda}\left(1450 {\rm \AA}\right)$\fi}
\newcommand{\fthree}{\ifmmode f_{\lambda}\left(3000{\rm \AA}\right) \else $f_{\lambda}\left(3000{\rm \AA}\right)$\fi}
\newcommand{\fH}{\ifmmode f_{\lambda}\left(1.65\micron\right) \else
$f_{\lambda}\left(1.65\micron\right)$\fi}

\newcommand{\fbol}{\ifmmode f_{\rm bol} \else $f_{\rm bol}$\fi}
\newcommand{\fbolwv}{\ifmmode f_{\rm bol}\left(\lambda\right) \else $f_{\rm bol}\left(\lambda\right)$\fi}
\newcommand{\fbolopt}{\ifmmode f_{\rm bol}\left(5100{\rm \AA}\right) \else $f_{\rm bol}\left(5100{\rm \AA}\right)$\fi}
\newcommand{\fbolthree}{\ifmmode f_{\rm bol}\left(3000{\rm \AA}\right) \else $f_{\rm bol}\left(3000{\rm \AA}\right)$\fi}
\newcommand{\fboluv}{\ifmmode f_{\rm bol}\left(1450{\rm \AA}\right) \else $f_{\rm bol}\left(1450{\rm \AA}\right)$\fi}

\newcommand{  \mbh      }{\ifmmode M_{\rm BH} \else $M_{\rm BH}$\fi}
\newcommand{  \lmbh     }{\ifmmode \log\left(\mbh/\Msun\right) \else $\log\left(\mbh/\Msun\right)$\fi} 
\newcommand{  \lledd    }{\ifmmode L/L_{\rm Edd} \else $L/L_{\rm Edd}$\fi}
\newcommand{  \Lbol     }{\ifmmode L_{\rm bol} \else $L_{\rm bol}$\fi}
\newcommand{  \lbol     }{\ifmmode L_{\rm bol} \else $L_{\rm bol}$\fi}
\newcommand{  \lLbol    }{\ifmmode \log\left(\Lbol/\ergs\right) \else $\log\left(\Lbol/\ergs\right)$\fi} 
\newcommand{  \Lagn     }{\ifmmode L_{\rm AGN} \else $L_{\rm AGN}$\fi}
\newcommand{  \lagn     }{\ifmmode L_{\rm AGN} \else $L_{\rm AGN}$\fi}

\newcommand{  \tgrow     }{\ifmmode t_{\rm growth} \else $t_{\rm growth}$\fi}
\newcommand{  \tUni      }{\ifmmode t_{\rm Universe} \else $t_{\rm Universe}$\fi}

\newcommand{  \Mindot	}{\ifmmode \dot{M}_{\rm infall} \else $\dot{M}_{\rm infall}$\fi}
\newcommand{  \Mbhdot	}{\ifmmode \dot{M}_{\rm BH} \else $\dot{M}_{\rm BH}$\fi}
\newcommand{  \Maddot	}{\ifmmode \dot{M}_{\rm AD} \else $\dot{M}_{\rm AD}$\fi}

\newcommand{  \as	}{\ifmmode a_{\rm *} 		\else $a_{\rm *}$\fi}
\newcommand{  \avec	}{\ifmmode \vec{a}_{\rm *} 	\else $\vec{a}_{\rm *}$\fi}
\newcommand{  \re	}{\ifmmode \eta      	\else $\eta$\fi}

\newcommand{  \mseed    }{\ifmmode M_{\rm seed} \else $M_{\rm seed}$\fi}
\newcommand{  \mbul     }{\ifmmode M_{\rm Bulge} \else $M_{\rm Bulge}$\fi} 
\newcommand{  \mstar    }{\ifmmode M_{*} \else $M_{*}$\fi} 
\newcommand{  \mgal     }{\ifmmode M_{*} \else $M_{*}$\fi} 
\newcommand{  \mhost    }{\ifmmode M_{\rm Host} \else $M_{\rm Host}$\fi}
\newcommand{  \mm       }{\ifmmode M_{*}/M_{\rm BH} \else $M_{*}/M_{\rm BH}$\fi}
\newcommand{  \mmsmall  }{\ifmmode M_{\rm BH}/M_{*} \else $M_{\rm BH}/M_{*}$\fi}
\newcommand{  \mmlarge  }{\ifmmode M_{*}/M_{\rm BH} \else $M_{*}/M_{\rm BH}$\fi}
\newcommand{  \mmwp     }{\ifmmode \left(M_{*}/M_{\rm BH}\right) \else $\left(M_{*}/M_{\rm BH}\right)$\fi}
\newcommand{  \ml       }{\ifmmode M_{*}/L_{*} \else $M_{*}/L_{*}$\fi}
\newcommand{  \mlwp     }{\ifmmode \left(M_{*}/L\right) \else $\left(M_{*}/L\right)$\fi}
\newcommand{  \mlk      }{\ifmmode \left(M_{*}/L_{K}\right) \else $\left(M_{*}/L_{K}\right)$\fi}
\newcommand{  \sigs     }{\ifmmode \sigma_{*} \else $\sigma_{*}$\fi}
\newcommand{  \Reff     }{\ifmmode R_{\rm e} \else $R_{\rm e}$\fi}

\def \swift {{\em Swift\ }}
\def \swiftxrt {{\em Swift} XRT\ }

\def \swiftbat {{\em Swift} BAT\ }
\def \swiftbatsh {{\em Swift} BAT}

\def \chandrash {{\em Chandra}}

\def \xmmsh{{\em XMM-Newton}}

\newcommand{\bj}{\ifmmode b_{\rm J} \else $b_{\rm J}$\fi}

\newcommand{\iab}{\ifmmode i_{\rm AB} \else $i_{\rm AB}$\fi}

\newcommand{\jab}{\ifmmode J_{\rm AB} \else $J_{\rm AB}$\fi}
\newcommand{\hab}{\ifmmode H_{\rm AB} \else $H_{\rm AB}$\fi}
\newcommand{\kab}{\ifmmode K_{\rm AB} \else $K_{\rm AB}$\fi}

\newcommand{\jveg}{\ifmmode J_{\rm Vega} \else $J_{\rm Vega}$\fi}
\newcommand{\hveg}{\ifmmode H_{\rm Vega} \else $H_{\rm Vega}$\fi}
\newcommand{\kveg}{\ifmmode K_{\rm Vega} \else $K_{\rm Vega}$\fi}

\def\arcsec{\hbox{$^{\prime\prime}$}}

\newcommand{  \Chisq    }{\ifmmode \chi^{2} \else $\chi^{2}$}
\newcommand{  \nelec    }{\ifmmode n_{e} \else $n_{e}$\fi}     
\newcommand{  \nh       }{\ifmmode n_{\rm H} \else $n_{\rm H}$\fi}     
\newcommand{  \Ncol     }{\ifmmode N_{col} \else $N_{col}$\fi} 
\newcommand{  \NH       }{\ifmmode N_{\rm H} \else $N_{\rm H}$\fi}     
\newcommand {\nhunit} {cm$^{-2}$}

\def\arcsec{\hbox{$^{\prime\prime}$}}
\def\ion#1#2{#1$\;${\small\rm\@Roman{#2}}\relax}

\newcommand{\OIIIa}{\ifmmode \left[{\rm O}\,\textsc{iii}\right]\,\lambda4959 \else [O\,{\sc iii}]\,$\lambda4959$\fi}
\newcommand{\NIIa}{\ifmmode \left[{\rm N}\,\textsc{ii}\right]\,\lambda6548 \else [N\,{\sc ii}]\,$\lambda6548$\fi}
\newcommand{\SIIa}{\ifmmode \left[{\rm S}\,\textsc{ii}\right]\,\lambda6716 \else [S\,{\sc ii}]\,$\lambda6716$\fi}
\newcommand{\SIIb}{\ifmmode \left[{\rm S}\,\textsc{ii}\right]\,\lambda6732 \else [S\,{\sc ii}]\,$\lambda6731$\fi}
\newcommand{\NeVa}{\ifmmode \left[{\rm Ne}\,\textsc{v}\right]\,\lambda3346 \else [Ne\,{\sc v}]\,$\lambda3346$\fi}
\newcommand{\NeVb}{\ifmmode \left[{\rm Ne}\,\textsc{v}\right]\,\lambda3426 \else [Ne\,{\sc v}]\,$\lambda3426$\fi}
\newcommand{\NeIIIa}{\ifmmode \left[{\rm Ne}\,\textsc{iii}\right]\,\lambda3869 \else [Ne\,{\sc iii}]\,$\lambda3869$\fi}
\newcommand{\NeIIIb}{\ifmmode \left[{\rm Ne}\,\textsc{iii}\right]\,\lambda3968 \else [Ne\,{\sc iii}]\,$\lambda3968$\fi}

\def\arcsec{{\mbox{$^{\prime \prime}$}}}

\def\ga{{\rm\thinspace gauss}}

\def\kpc{{\rm\thinspace kpc}}
\def\Lsun{\hbox{$\rm\thinspace L_{\odot}$}}

\def\Msun{\hbox{$\rm\thinspace M_{\odot}$}}
\def\pc{{\rm\thinspace pc}}

\newcommand{  \Ntot    }{559}
\newcommand{  \Ntott    }{340}

\title[BASS II: X-rays vs optical emission lines]{BAT AGN spectroscopic survey II: X-ray emission and high ionization optical emission lines}
\author[Berney et al.]
{\parbox{\textwidth}{Simon Berney$^{1}$, Michael Koss$^{1}$\thanks{E-mail: mike.koss@phys.ethz.ch}\thanks{Ambizione fellow}, Benny Trakhtenbrot$^{1}$\thanks{Zwicky Fellow}, Claudio Ricci$^{2}$, Isabella Lamperti$^{1}$, Kevin Schawinski$^{1}$, Mislav Balokovi{\' c}$^{3}$, D. Michael Crenshaw$^{4}$, Travis Fischer$^{4}$, Neil Gehrels$^{5}$, Fiona Harrison$^{3}$, Yasuhiro Hashimoto$^{6}$, Kohei Ichikawa$^{7}$, Richard Mushotzky$^{8}$, Kyuseok Oh$^{1}$, Daniel Stern$^{9}$, Ezequiel Treister$^{10}$, Yoshihiro Ueda$^{7}$, Sylvain Veilleux$^{8}$, and Lisa Winter$^{11}$}\vspace{0.4cm}\\
\parbox{\textwidth}{
$^{1}$Institute for Astronomy, Department of Physics, ETH Zurich, Wolfgang-Pauli-Strasse 27, CH-8093 Zurich, Switzerland\\
$^{2}$Instituto de Astrof\'{\i}sica, Facultad de F\'{\i}sica, Pontificia Universidad Cat\'olica de Chile, Casilla 306, Santiago 22, Chile\\
$^{3}$Cahill Center for Astronomy and Astrophysics, California Institute of Technology, Pasadena, CA 91125, USA\\
$^{4}$Department of Physics and Astronomy, Georgia State University, Astronomy Offices, One Park Place South SE, Suite 700, Atlanta, GA 30303, USA\\
$^{5}$NASA Goddard Space Flight Center, Greenbelt, MD 20771, USA\\
$^{6}$Department of Earth Sciences, National Taiwan Normal University, No. 88, Sec. 4, Tingzhou Rd., Wenshan District, Taipei
11677, Taiwan R.O.C\\
$^{7}$Department of Astronomy, Kyoto University, Kyoto 606-8502,
Japan\\
$^{8}$Astronomy Department, University of Maryland, College Park, MD, USA\\
$^{9}$Jet Propulsion Laboratory, California Institute of Technology, 4800 Oak Grove Drive, MS 169-21, Pasadena, CA 91109, USA\\
$^{10}$Departamento de Astronom\'{\i}a, Universidad de Concepci\'{o}n, Concepci\'{o}n, Chile\\
$^{11}$Atmospheric and Environmental Research, 131 Hartwell Ave No. 4, Lexington, MA 02421, USA\\
$^{\dagger}$Ambizione fellow\\ 
$^{\ddagger}$Zwicky fellow
}}

\begin{document}

\date{
}
\pagerange{\pageref{firstpage}--\pageref{lastpage}} \pubyear{2015}

\maketitle

\label{firstpage}

\begin{abstract}
We investigate the relationship between X-ray and optical line emission in \Ntott\ nearby ($z\simeq 0.04$) AGN selected above 10 \kev\ using \swiftbatsh.
We find a weak correlation between the extinction corrected \oiii\ and hard X-ray luminosity ($L_{\oiii}^{int} \propto L_{14-195}$) with a large  scatter ($\rm{R_{Pear} = 0.64}$, $\rm{\sigma=0.62}$ dex) and a similarly large scatter with the intrinsic 2$-$10 keV to \oiii\ luminosities ($\rm{R_{Pear} = 0.63}$, $\rm{\sigma=0.63}$ dex).  Correlations of the hard X-ray fluxes with the fluxes of high-ionization narrow lines (\oiii, \heii, \neiii\ and \nev) are not significantly better than with the low ionization lines (\ha, \sii). Factors like obscuration or physical slit size are not found to be a significant part of the large scatter. In contrast, the optical emission lines show much better correlations with each other ($\sigma=0.3$ dex) than with the X-ray flux. The inherent large scatter questions the common usage of narrow emission lines as AGN bolometric luminosity indicators and suggests that other issues such as geometrical differences in the scattering of the ionized gas or long term AGN variability are important.

\end{abstract}

\section{Introduction}

Based on the unification model of active galactic nuclei  \citep[AGN;][]{urry91, antonucci93, urry95}, narrow optical emission lines are emitted from the narrow-line region (NLR). The NLR is ionized by the Extreme UV--soft X-ray emission from the AGN's central engine.  Afterwards, the illuminated interstellar gas in the NLR radiatively cools down by emitting the narrow optical lines. Unlike this optical emission, X-rays are emitted from much closer to the supermassive black hole (SMBH). The UV photons radiated by the accretion disk are inverse Compton scattered inside the corona and can reach energies up to a few hundreds of \kev\ \citep[e.g.,][]{netzer}. Thus, if AGN are accreting at a constant rate and constantly ionizing the NLR, we would naively expect a tight correlation between the X-ray emission from the corona and the optical emission lines from the NLR. This is why both high ionization optical emission lines and X-rays are thought to be reliable tracers of the AGN bolometric luminosity ($L_{bol}$) and of the black hole growth \citep{heck05}.\\


The bolometric luminosity ($L_{bol}$) of unobscured AGN estimated from  the spectral energy distribution (SED) fitting technique using optical, UV, and X-ray data
shows a tight correlation with hard X-ray data alone \citep[e.g., 14$-$195 \kev\ with a scatter of 0.3 dex,][]{vasudevan} though X-ray surveys may not detect heavily obscured Compton-thick AGN ($\NH>10^{24}$ \nhunit).
Strong observational correlation has also been found between UV emission and X-ray emission \citep{steffen06, young10}.
The use of \OIII\ to infer AGN bolometric luminosities has a long history \citep[e.g.,][]{heckman2004, lamassa, heckman2009}. It has been widely used to estimate the accretion rates of AGN using bolometric corrections based on the X-ray emission, the 5100 \AA\ monochromatic continuum or the mid-IR emission \citep[e.g.,][]{oiiibol}. However, several studies have noted that the strength of \OIII\ is modulated by factors including covering factor of the narrow-line region gas and its density \citep[e.g.,][]{netzer93}, and ionization parameter \citep[e.g.,][]{baskin2005}.\\

Previous studies have compared X-ray emission with optical emission lines strengths in hard X-ray selected samples.
\cite{heck05} used a sample of 47 local AGN ($z<$0.2) from the \textit{RXTE} slew survey to analyze the correlation between 3-20 \kev\ flux and the \oiii\ strength \citep[][]{rxte, rxte2}.
Other works have also made use of the early results from the \swiftbat survey \cite[9-month catalog,][]{tueller} to analyze optical and X-ray emission from AGN.
\cite{melendez} studied the correlation between the \OIII\ line strength and two X-ray bands (2$-$10 \kev\ and 14$-$195 \kev) for 40 low redshift (z$<$0.08) BAT AGN from the 9-month catalog. 
\cite{lwinter} made a similar analysis with  a slightly larger sample (64 BAT AGN). Finally, an analysis of the full 9-month catalog has recently been completed ($\approx$100 BAT AGN, Ueda et al. 2015, submitted). \cite{netzer2006} studied a sample of 52 obscured AGN at high redshift ($z\approx$1) suggesting that \oiii\ emission is weaker relative to 2$-$10 \kev\ emission in high $L_X$ sources.\\

The \swiftbat 70-month all-sky survey enables us to analyze the optical and X-ray properties of a large sample of nearby AGN. It has detected 823 AGN to a flux limit of $\approx 10^{-11}\ \mathrm{erg\ cm^{-2}\ s^{-1}}$ over the whole sky \citep{Baumgartner2012aa}.
Due to its detection method in the ultra hard X-rays (14$-$195 \kev), the \swiftbat survey is not affected by obscuration (below Compton-thick levels) nor is it contaminated by stellar emission. Therefore, it also includes 'buried' obscured AGN that may not have been selected by optical surveys  \citep[e.g.,][]{ueda}.\\

In this work, we use a sample of BAT AGN based on the 70-month catalog with an optical spectroscopic sample that is significantly larger than previous studies (\Ntott\ objects). Our spectroscopic analysis also includes higher ionization lines not used previously (i.e., \NeV) to study correlations between optical line emission and X-ray emission.  The sample is presented in Section 2 and our analysis methods are described in Section 3.  We study the correlation between \oiii\ and X-ray emission in Section 4.1, X-ray and other high ionization lines (i.e., \NeIIIa, \HeIIop, and \NeV) in Section 4.2, and the correlation between only optical emission lines in Section 4.3.  Finally, a review of our results and a discussion is presented in Section 5. \\
\section[]{Sample}
The optical data are taken from the BAT AGN Spectroscopic Survey (BASS) Data Release 1 (Koss et al., in prep.), which gathered \Ntot\ optical spectra from targeted spectroscopic campaigns on BAT sources and public optical surveys \citep[SDSS and 6DF,][]{sdss,6df}.
This sample is the largest and the most complete catalog of AGN optical spectra selected from the \swift\ hard X-ray survey. It contains 67.6\% of the total AGN detections in the \swiftbat 70-month catalog and has a mean redshift of $z=0.10$.\\

From this parent sample, we first remove the spectra coming from the 6dF survey because they lack flux calibration.  We remove all blazars listed in the Roma blazar catalog (BZCAT) v5.0 \citep{bzcat} since the X-ray emission in such sources is affected by differential beaming. We also apply two redshift cuts. The first one, $z>0.01$, avoids very nearby star-forming galaxies (such as M82) and ensures that a significantly large portion of the ionized region within the galaxy is covered (i.e., $>$0.2 \kpc). We also limit to $z<0.4$ to avoid redder emission lines such as \ha\ from being shifted outside the observed wavelength interval. Additionally, we remove ten spectra due to calibration problems around the \oiii\ line. Finally, we exclude SDSS J155334.73+261441.4, a gravitationally lensed quasar \citep{lensed}.
The final sample totals \Ntott\ optical spectra with $\langle z \rangle =0.05$.\\


We define an emission line detection when we reach a SNR$>$3 of the line with respect to the noise of the adjacent continuum. Out of these \Ntott\ objects, we have 327 detections of \oiii, 301 of \ha, 281 of \nii, 278 of \sii, 265 of \hb, 206 of \oi, 144 of \neiii, 106 of \heii\ and 51 of \nev. The intrinsic 2$-$10 \kev\ flux values and the column density measurements provided for 335/\Ntott\ AGN are based on a homogeneous model fitting using the best available X-ray data with simultaneous modeling of the 0.2$-$10 keV band from \xmmsh, \chandrash, or \swiftxrt and the 14$-$195 keV band from \swiftbat  (details in Ricci et al., in prep).\\

We use the AGN emission-line diagnostics \citep[e.g.,][]{BPT, veilleux} as updated by \citet{kew06} to further classify our sample.  We classify as low ionization nuclear emission-line region (LINER) any object that shows a LINER classification in either the \OI\ or \SII\ diagnostic. For the remaining AGN, we use the \NII\ diagnostic, which is most sensitive to AGN, to differentiate between Seyfert AGN and HII/composite/ambiguous (17/\Ntott).  We use the presence of a broad line in \Halpha\ to differentiate between type 1 (217/\Ntott) and type 2 objects (116/\Ntott).  A more detailed description of the reduction procedures of the optical spectra, sample selection, and overall properties of the optical spectroscopic sample can be found in the BASS-I paper (Koss et al., in prep).
\section[]{Analysis of optical spectra}
We fit our sample of optical spectra using an extensive spectroscopic analysis toolkit for astronomy, {\tt PySpecKit}, which uses a Levenberg-Marquardt algorithm for fitting \citep{pyspeckit}.  We implement separate methods for fitting sources with narrow lines only and sources with broad lines.  We fit the \hb\ (4650-5050 \AA), \ha\ (6250-6770 \AA), and \oii\ (3300-4000 \AA) spectral regions separately.  For narrow-line sources, we first fit the spectrum with stellar templates using the {\tt pPXF} software \citep[Penalized PiXel Fitting;][]{ppxf} to remove the galaxy continuum and stellar absorption features that can effect line measurements.  All emission line fits were inspected by eye to verify proper fitting.\\  

We adopt a power-law fit (1st order) to model the continuum and Gaussian components to model the emission lines for different spectral regions.  For the \Halpha\ spectral region, we fit the \OI, \NIIa, \Halpha, \NII, and \SIIa, and \SIIb\ lines. The narrow-line widths are tied together with an allowed variation of $\pm$ 8 \AA\ and the systemic redshift is determined from the \Halpha\ line. The relative strengths of \NIIa\ and \NII\ lines are fixed (at 1:2.94).  For the \Hbeta\ spectral region, we fit the \HeIIop, \Hbeta, \OIIIa, and \OIII\ lines. The narrow-line widths are also tied together with an allowed variation of $\pm$ 8 \AA\ and their central wavelengths are defined by the redshift of \OIII. The intensity of \OIIIa\ relative to \OIII\ is fixed (at 1:2.86).  For the \oii\ spectral region, we fit the \NeVa, \NeVb, \OII, \NeIIIa, and \NeIIIb\ lines.\\

Two Gaussian components are allowed for the \Halpha\ and \Hbeta\ emission lines: a narrow component (FWHM $< 1000\  \kms$) and a broad component (FWHM $> 1000\ \kms$). The broad \Hbeta\ component is only allowed if a broad \Halpha\ component is detected. In the case of a non-detection of the narrow \Halpha\ or \SII\ line (i.e., due to a very strong broad \Halpha\ component, a noisy spectrum or bad sky subtraction), we use the FWHM of \OIII\ to constrain the widths of the narrow lines in the \Halpha\ region.\\
 
\begin{figure*}
\centering
\subfigure{\includegraphics[width=0.49\textwidth]{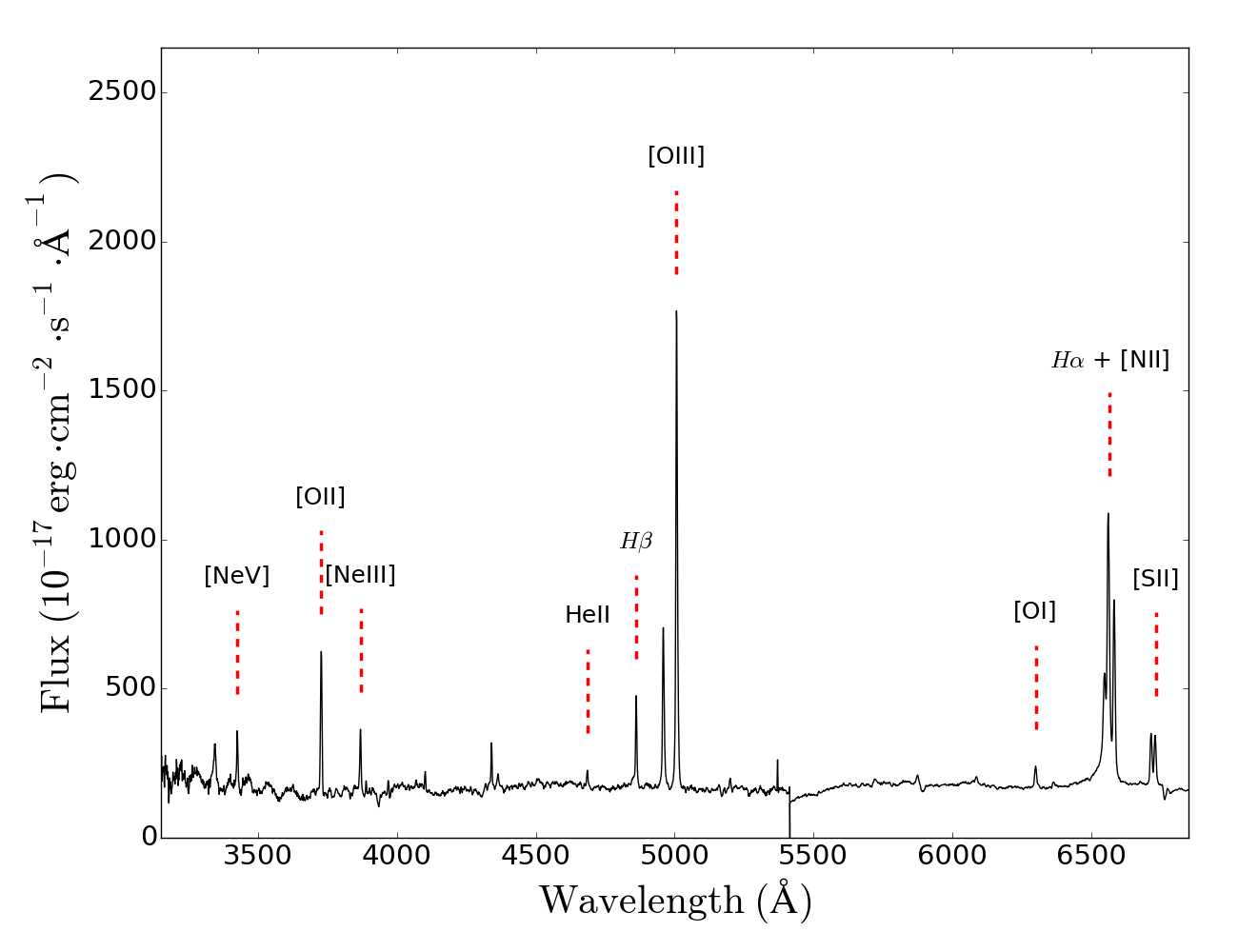}}
\hfill
\subfigure{\includegraphics[width=0.49\textwidth]{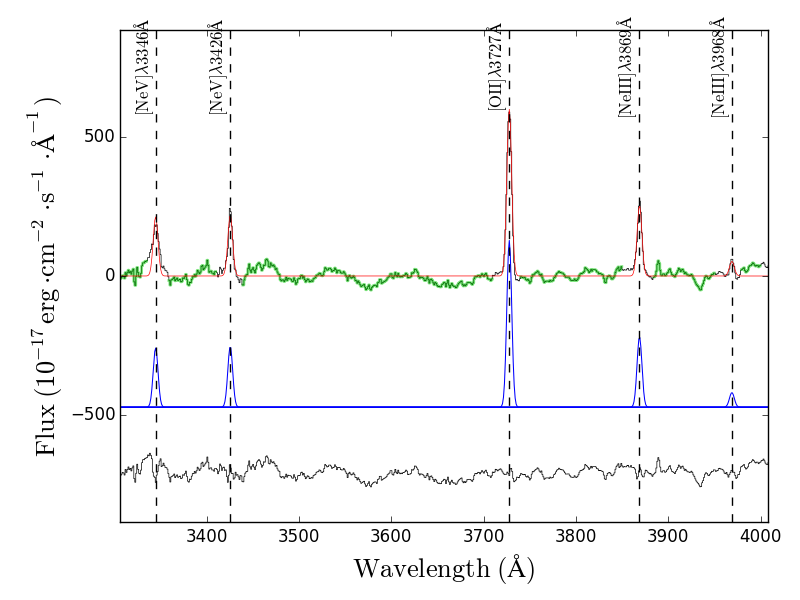}}
\vfill
\subfigure{\includegraphics[width=0.49\textwidth]{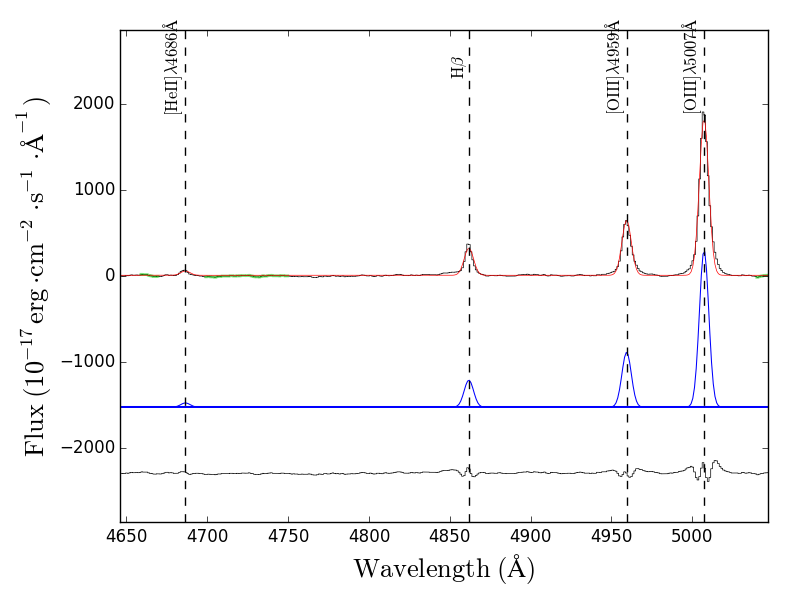}}
\hfill
\subfigure{\includegraphics[width=0.49\textwidth]{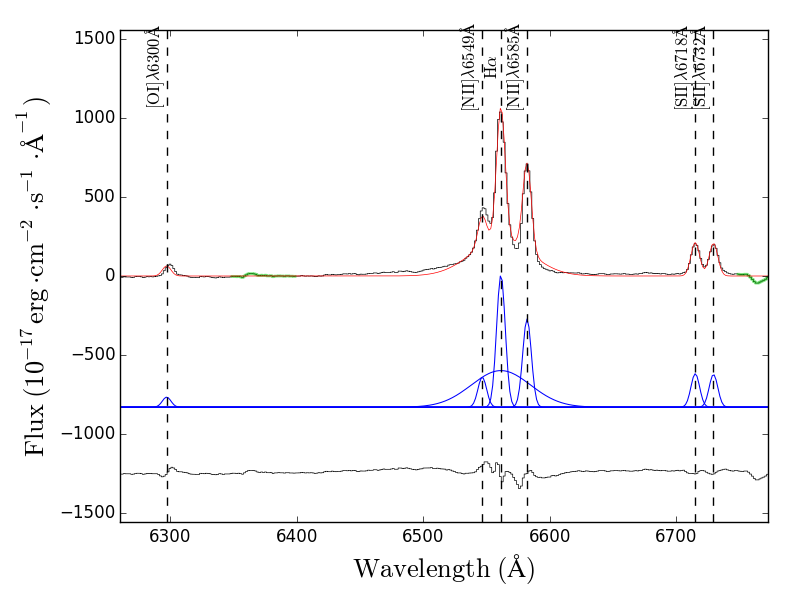}}

\caption{Example of the fitting procedure for MCG +06-16-28 observed with the Palomar Double Spectrograph.
\textit{Top left panel}: Spectrum of the object with all the fitted emission lines. Rest frame continuum subtracted fits are shown for the \oii\ (top right), \hb\ (bottom left), and \ha\ (bottom right) spectral regions.
The colors denote regions used for continuum (green), different Gaussian components of the fit (blue), and model fits (red) with residuals shown in the bottom panel.}
\label{example_fit}
\end{figure*}

To estimate the continuum for the \Halpha\ complex, we use  the wavelength regions 5800-6250 \AA\ and 6750-7000 \AA.
For the \Hbeta\ complex, we use the 4660-4750 \AA\ (except around \heii) and 5040-5200 \AA\ regions. The continuum around the \oii\ spectral region is usually more complicated to fit due to a non-linear shape or because it lies at the blue limit of the wavelength coverage. To fit this blue continuum, we use the region between 3300 \AA\ and 4000 \AA\ except where the emission lines are located ($\pm$ 15 \AA). To estimate errors in line fluxes and line widths, we use a Monte Carlo simulation that adds noise based on the error spectrum and reruns the fitting procedure 10 times. The flux uncertainty for the \oiii\ emission line is typically less than 1\%.
To correct our line ratios for dust extinction, we use the narrow Balmer line ratio (\ha/\hb) assuming an intrinsic ratio of R=3.1 \citep[e.g.,][]{ferland1986} and the \citet{cardelli} reddening curve when both the narrow \ha\ and \hb\ are detected. Figure \ref{example_fit} shows an example spectrum and best model fit.\\
\section[]{Results}
\subsection[]{X-ray and \OIII\ Flux}

\begin{figure*} 
\centering
\includegraphics[width=0.89\textwidth]{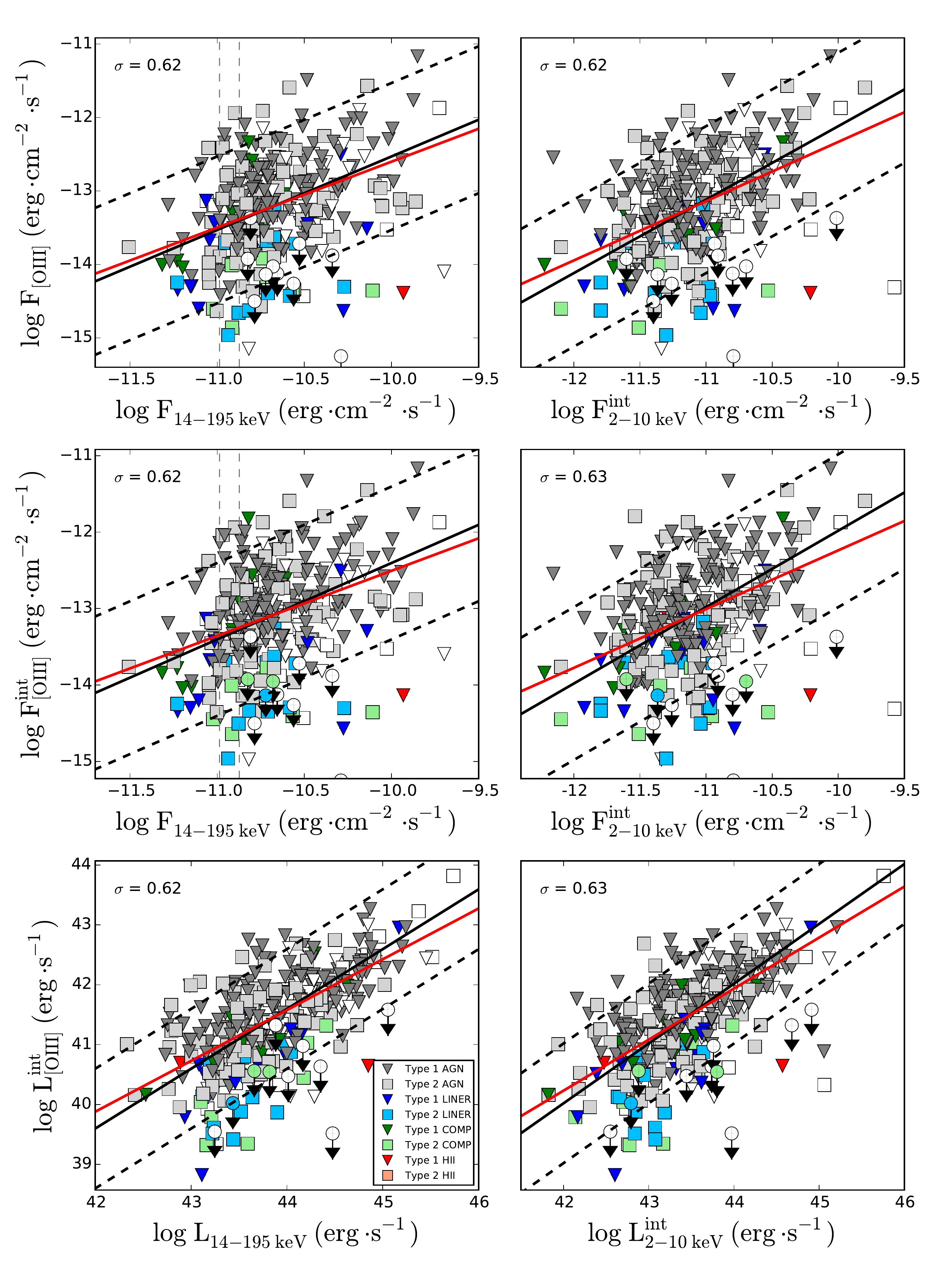}
\caption{Plots of the relation of the \OIII\ emission line strength with the BAT X-ray emission (\textit{left}) and with the 2$-$10 \kev\ emission (\textit{right}). The \oiii« -- »X-ray relation exhibits a large scatter in all panels ($\approx$0.6 dex). The black solid lines show a fit constrained by $Y \propto X$ with the dashed lines showing the $\pm$1 dex. The red lines illustrate the linear regression. For both line fits we removed the four largest outliers. The color code comes from the the \nii\ BPT classification. White symbols do not have any BPT classification because of undetected emission lines. The AGN showing broad \ha\ emission are classified as type 1. On the top and central left panels the grey dashed lines illustrate the 14$-$195 \kev\ flux limits for 50\% (left line) and 90\% of the sky (right line). The upper limits were ignored in the analysis.
\textit{Top panels}: observed \oiii\ flux plotted against the X-ray fluxes. 
\textit{Central panels}: extinction corrected \oiii\ flux plotted against the X-ray fluxes. \textit{Lower panels}: extinction corrected \oiii\ luminosity vs. the two X-ray luminosities.}

\label{flux_OIII}
\end{figure*}

We first compare the X-ray flux (2$-$10 \kev\ and 14$-$195 \kev) to the \OIII\ emission line (Figure \ref{flux_OIII}).
The relation between the \oiii\ corrected flux and the BAT X-ray flux shows a scatter of $\sigma$=0.62 dex and a Pearson correlation coefficient of $\rm{R_{Pear}=0.35}$. The result is very similar when comparing \oiii\ to the 2$-$10 \kev\ flux ($\sigma$=0.63, $\rm{R_{Pear}=0.38}$). The Pearson $p$-value representing the probability of the hypothesis to have no correlation at all between the two X-ray fluxes and the \oiii\ flux is $<10^{-10}$ because of the large sample size. The correlation in luminosity space is better ($\rm{R_{Pear}^{14-195}=0.64}$, $\rm{R_{Pear}^{2-10}=0.63}$) than in flux space, as expected because the same multiplicative factor ($4\pi d_{L}^2$) is applied to both the X-ray flux and the \oiii\ flux to obtain the luminosities\footnote{$d_{L}$ varies here from 43 Mpc to 2.2 Gpc}. We finally compare the \OIII\ flux both with ($F_{\oiii}^{int}$) and without ($F_{\oiii}$) extinction correction to the two X-ray band fluxes. We find no significant improvement in the scatter of the different relations ($\sim$0.01 dex) using the Balmer decrement corrected \oiii\ fluxes.\\

Next, we compare a one-to-one relation and the linear regression in logarithmic space to fit our data. For both slope fits, we remove the four largest outliers. For the line fitting, we use the python routine {\tt scipy.optimize.curve\_fit} which provides a least squares approach (OLS(Y$\mid$ X)) with the Levenberg-Marquardt gradient method for convergence.
The black solid lines in the panels of Figure \ref{flux_OIII} show the best one-to-one relation (which would imply that $L_{\oiii}^{int} \propto L_{14-195}$). The best fit for the linear regression, shown with the red lines, is $L_{\oiii}^{int} \propto (L_{14-195})^{0.85\pm0.05}$ and $L_{\oiii}^{int} \propto (L_{2-10}^{int})^{0.85\pm 0.05}$.
To test if the linear regression is a significantly better fit, we run a $F$-test. The $F$-test compares the sum of the residuals squared of two different models. We obtain a $F$-value of 7.66 for the 14$-$195 \kev\ flux and of 7.81 for the 2$-$10 \kev\ flux. The resulting $p$-values are lower than 1\% suggesting that the linear regression between the \oiii\ luminosity and the two X-ray bands is a significantly better model than a one-to-one relation. However, the improvement in the scatter is very small ($\approx$ 0.01 dex). Therefore, we decided to use the one-to-one relation (the simplest model)  for the rest of the analysis.\\  

We also use the {\it Astronomy Survival Analysis} ({\sc ASURV}) package from the {\sc STATCODES} suite of utilities developed by Eric Feigelson (Feigelson 1985). We determine the correlation between \oiii\ and X-ray luminosity in the presence of the 13 upper limits in \oiii.  Specifically, we use the EM algorithm, though the Buckley James yields similar results. We find the best fit correlations is $L_{\oiii}^{int} \propto (L_{14-195})^{0.83\pm0.06}$ which is consistent with the least squares approach suggesting the small number of non-detections do not significantly change the slope of the correlation.\\

Finally, we use the OLS bisector method of line fitting which is more reliable at recovering the intrinsic slope in the presence of uncertainties in both the X and Y variables (e.g. Isobe 1990).  Table \ref{tablelum} lists the best fit linear regressions for the luminosity-luminosity plots using the OLS bisector method as well as the standard deviation for the complete sample and various sub-samples (Type 1-1.9, Type 2, Seyferts, LINERs, and HII/composite/ambiguous).  The LINERS and HII/composite objects \citep[from the BPT classification;][]{BPT, veilleux} have on average lower X-ray and \oiii\ luminosities than Seyfert galaxies:  $\langle \log\ L_{\oiii, int}^{LIN}\rangle=40.5$; $\langle \log\ L_{\oiii, int}^{comp}\rangle= 40.9$; $\langle\log\ L_{\oiii, int}^{Sey}\rangle=41.5$. \\


\subsubsection[]{Correlations with obscuration and luminosity}
\begin{figure*} 
\centering
\includegraphics[width=0.89\textwidth]{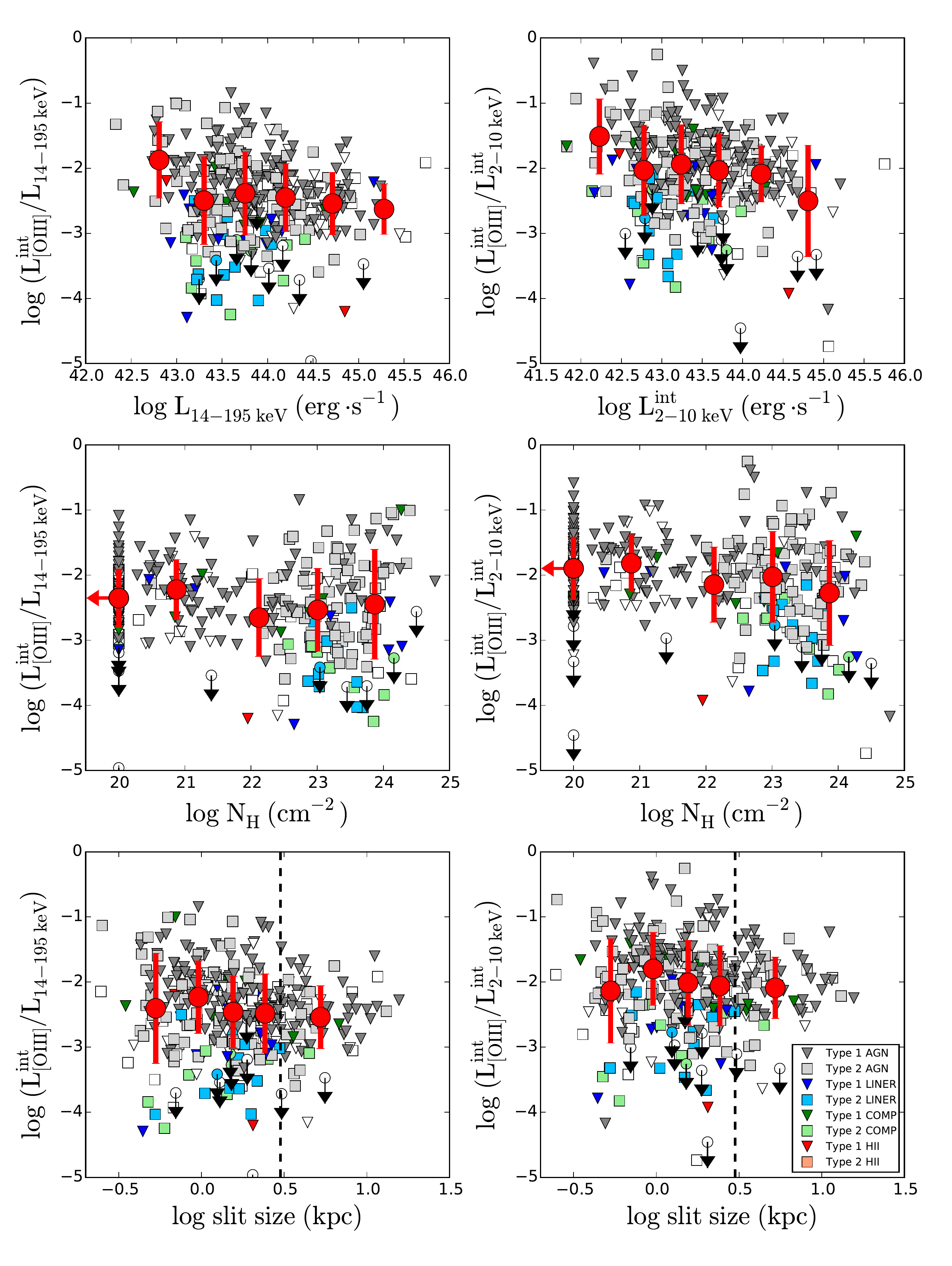}
\caption{Plots of the $L_{\oiii}^{int}/L_{X}$ ratio as a function of X-ray luminosity (\textit{top panels}), column density (\textit{central panels}), and physical slit size (\textit{lower panels}). The color codes represent the \nii\ BPT classification. White symbols do not have any BPT classification because of undetected emission lines. The red dots represent average binned values and the red bars show 1-sigma deviations in the bins. The first bin of the central panels represents a upper limit where the column density is consistent with the Galactic neutral hydrogen.  The horizontal black lines on the lower panels illustrate the $3\ kpc$ value. The scatter increases for $\NH>10^{22}$ \nhunit\ with no change in average $L_{\oiii}^{int}/L_{X}$.  For slit size, the scatter decreases with increasing slit size because of a higher fraction of unobscured AGN at higher redshifts, but the average $L_{\oiii}^{int}/L_{X}$ is constant. The upper limits were ignored in the analysis.}
\label{nh}
\end{figure*}

\begin{figure*} 
\centering
\includegraphics[width=0.89\textwidth]{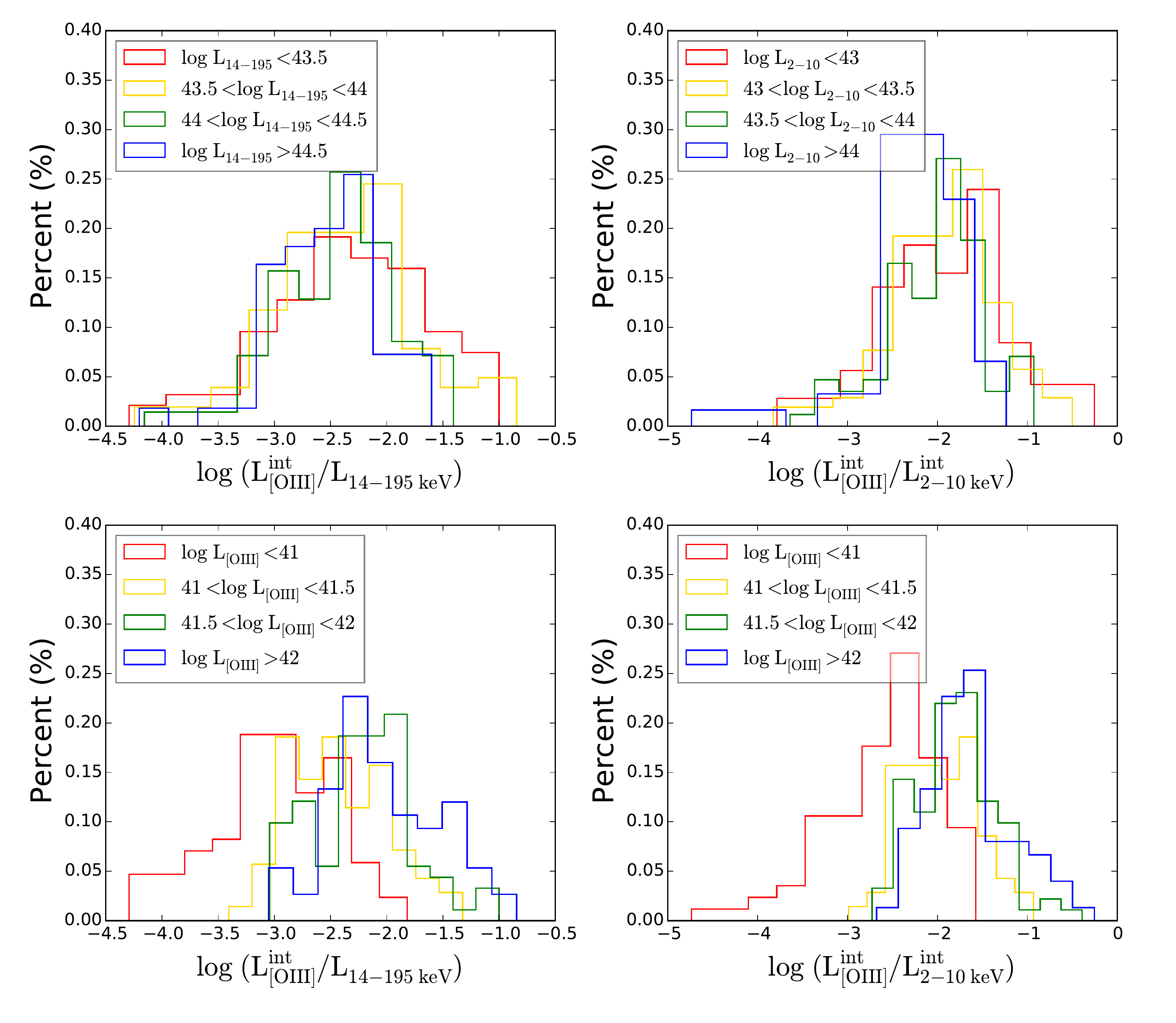}
\caption{Histograms of the $L_{\oiii}^{int}/L_{X}$ ratio binned by X-ray luminosity (\textit{top panels}) and \oiii\ luminosity (\textit{lower panels}). The two lower panels show a strong positive correlation of the ratio with the \oiii\ luminosity.}
\label{lum_bin}
\end{figure*}

The two upper panels of Figure \ref{nh} show the $L_{\oiii}^{int}/L_{X}$ ratio as a function of X-ray luminosity. The red dots show the average trend and the error bars the average 1-$\sigma$ scatter.
First, we notice that there is no significant difference between the 2$-$10 \kev\ (right panels) and the BAT (left panels) X-ray luminosity relations. There is a slight decrease of the \oiii/X-ray flux ratio when we go to higher X-ray luminosities, which is consistent though with being constant at a 3-$\sigma$ level.  Figure \ref{lum_bin} illustrates the luminosity dependence of the flux ratio using histograms.  The scatter slightly decreases towards high X-ray luminosity. This trend is not statistically significant, and is probably driven by the increase of the unobscured AGN fraction with high luminosities \citep[receding torus model, e.g.,][]{lawrence91, simpson, simpson05, brightman15} since unobscured AGN exhibit lower scatter than the obscured AGN (see central panels of Figure \ref{nh}, see below). The two lower panels of Figure \ref{lum_bin} show a strong increase of the flux ratio as a function of the \oiii\ luminosity. \\

Dust and obscuration might still play a significant role in the relation between optical and X-ray emission even if the optical emission lines are corrected for host galaxy extinction using the Balmer decrement. The obscuring torus scatters a portion of the ionizing light from the nucleus and therefore also has an indirect effect on the optical emission line intensities.
The two central panels of Figure \ref{nh} show the $L_{\oiii}^{int}/L_{X}$ ratio as a function of the integrated  column density \NH. Our results show that the trend of the luminosity ratio is consistent with being constant from very low to very high levels of obscuration. This is true for the intrinsic 2$-$10 \kev\ emission as well as for the 14$-$195 \kev\ band. We obtain average ratios of $\log\ L_{\oiii}^{int}/L_{14-195}=-2.42$ and $\log\ L_{\oiii}^{int}/L_{2-10}^{int}=-2.01$. However, the scatter strongly increases for \NH$>10^{22}$ \nhunit.
We observe a scatter of $\sigma=0.47$ for low obscuration AGN (\NH$<10^{22}$ \nhunit) and of $\sigma=0.72$ for high obscuration AGN (\NH$>10^{22}$ \nhunit) in the case of the BAT X-ray band. The result is similar for the 2$-$10 \kev\ emission.   In order to quantify this change in scatter, we run a Levene's test.  This test checks whether the two samples are drawn from distributions having a similar variance.  We obtain a $p$-value = $6.52\times 10^{-6}$ for the BAT band and of $2.35\times 10^{-4}$ for the 2$-$10 \kev\ band, implying a highly significant increase in the scatter from low to high obscuration in both X-ray bands.\\

\subsubsection[]{Slit Size and Dilution}
Due to different slit widths (0.75-3.0\arcsec) used to observe the BAT AGN and the various redshifts of these AGN (0.01$<z<$0.4), our spectroscopic data cover regions between 0.2 \kpc\ to the size of the entire host galaxy. In addition to that, the NLR size may vary from one galaxy to another and the variation is strongly dependent on the power of the AGN \citep[e.g.,][]{mor2009, nlr}.  This change in the size of the observed region and in the size of the NLR has various implications for our measurements.  First, the region we probe with the spectroscopic observations might not cover the entire ionized \oiii\ region for luminous sources at low redshift. We might therefore underestimate the \oiii\ flux for these objects. Alternatively in low luminosity AGN, the region we probe with the slit could be much larger than the size of the NLR with a significant contaminating component from the host galaxy.\\


We study $L_{\oiii}^{int}/L_{X}$ as a function of the physical slit size in kpc (lower panels of Figure \ref{nh}) to study these issues. Our results show that the scatter slightly decreases towards larger slit sizes while the ratio $L_{\oiii}^{int}/L_{X}$ is constant.  To further quantify this trend, we split our sample in two based on slit sizes $>3$ \kpc\ and slit sizes $<3$ \kpc.  We observe that the scatter decreases by respectively 0.16 dex and 0.20 dex for the 14$-$195 \kev\ and 2$-$10 \kev\ band.  However, this decrease in scatter is probably driven by the increase of the unobscured AGN fraction with high redshift rather than simply larger physical slit sizes.  

\subsubsection[]{Sample inhomogeneity}

\begin{figure*}
\centering
\includegraphics[width=0.95\textwidth]{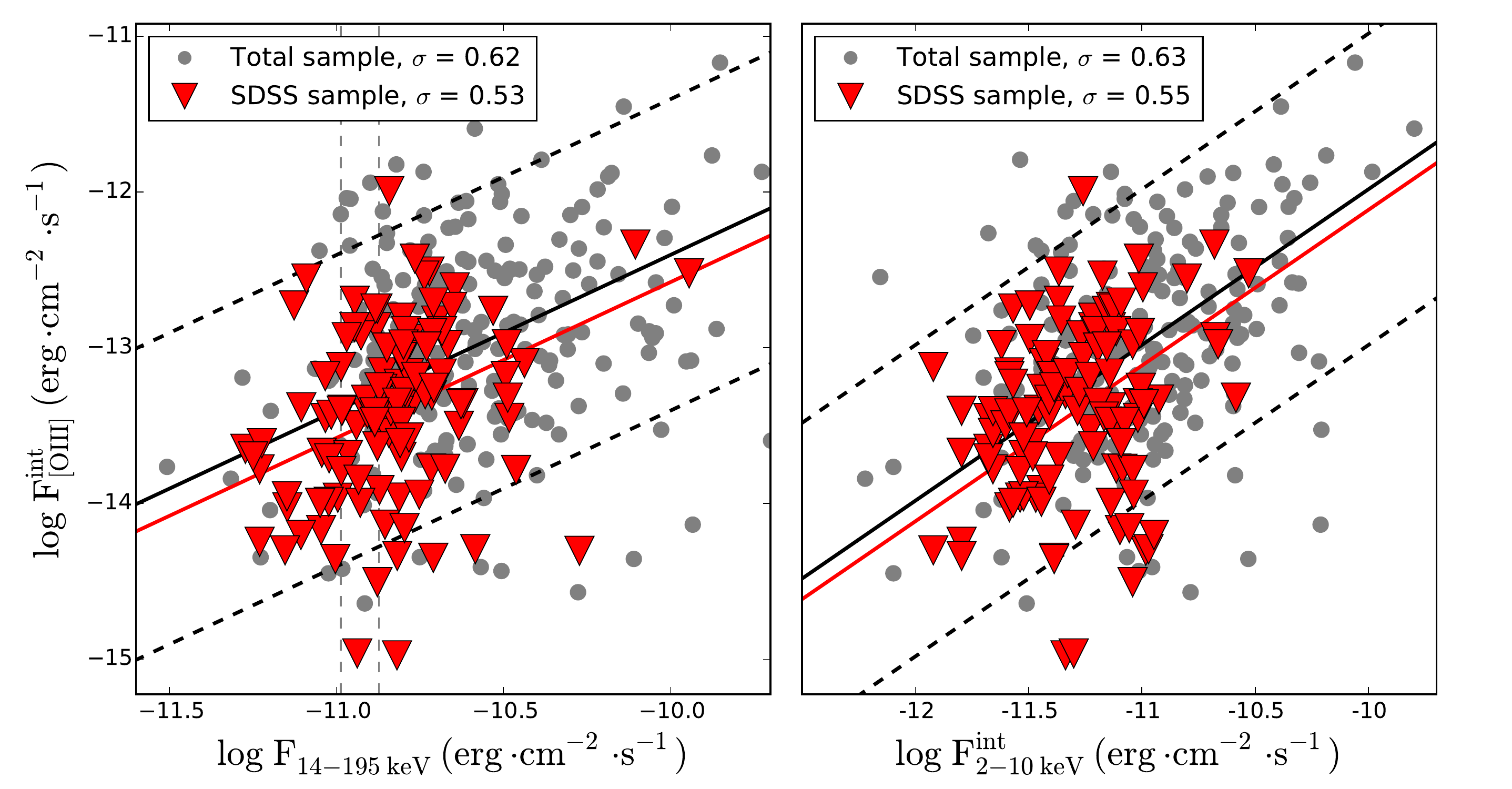}
\caption{Plots of the \OIII\ extinction corrected flux vs the two X-ray fluxes. The red triangles represent the sources observed by SDSS (110 objects). The black line is a one-to-one fit of the total sample and the red line a one-to-one fit of the SDSS sub-sample. The scatter and the fit of the sub-sample are comparable to the ones of the main sample.}

\label{sdss}
\end{figure*}

The BASS survey gathered optical spectra from various instruments with different data qualities (i.e., resolution, sensitivity, wavelength range). This lack of homogeneity in the data could artificially increase the scatter in the \oiii« -- »X-ray relation. For this reason, we also analyze the SDSS sub-sample (110 AGN) because of their high quality in flux calibration (Figure \ref{sdss}). We find that the scatter decreases by 0.09 dex for the 14$-$195 \kev\ and 0.08 dex for the 2$-$10 \kev\ band. In order to have a quantitative result for this change in scatter, we run again a Levene's test. The resulting $p$-values for the SDSS sub-sample are respectively 0.062 and 0.35  for the BAT and 2$-$10 keV X-ray bands which is not statistically significant.
The fit of the SDSS sub-sample (red line) is similar to the fit of the total sample (black line) for both X-ray bands. This implies that we do not have any statistical offset due to systematics (like flux calibration issue). We conclude that there is no significant difference between the total sample and the SDSS sub-sample.

\subsection[]{X-ray correlation with high and low ionization lines}


Lines produced from high ionization ($>$40 eV) have the advantage of being less contaminated by host galaxy emission than lines produced from low ionization ($<$20 eV). Moreover, these lines are emitted in less extended regions, which implies that we cover a larger portion of their emitting regions in our observations than for low ionization lines. Finally, we include the \Halpha\ and \Hbeta\ emission lines as these lines are either not or less affected with respect to other lines by variations in metal abundances, ionization levels, and  collisional effects.   \\

\begin{figure*} 
\centering
\includegraphics[width=0.85\textwidth]{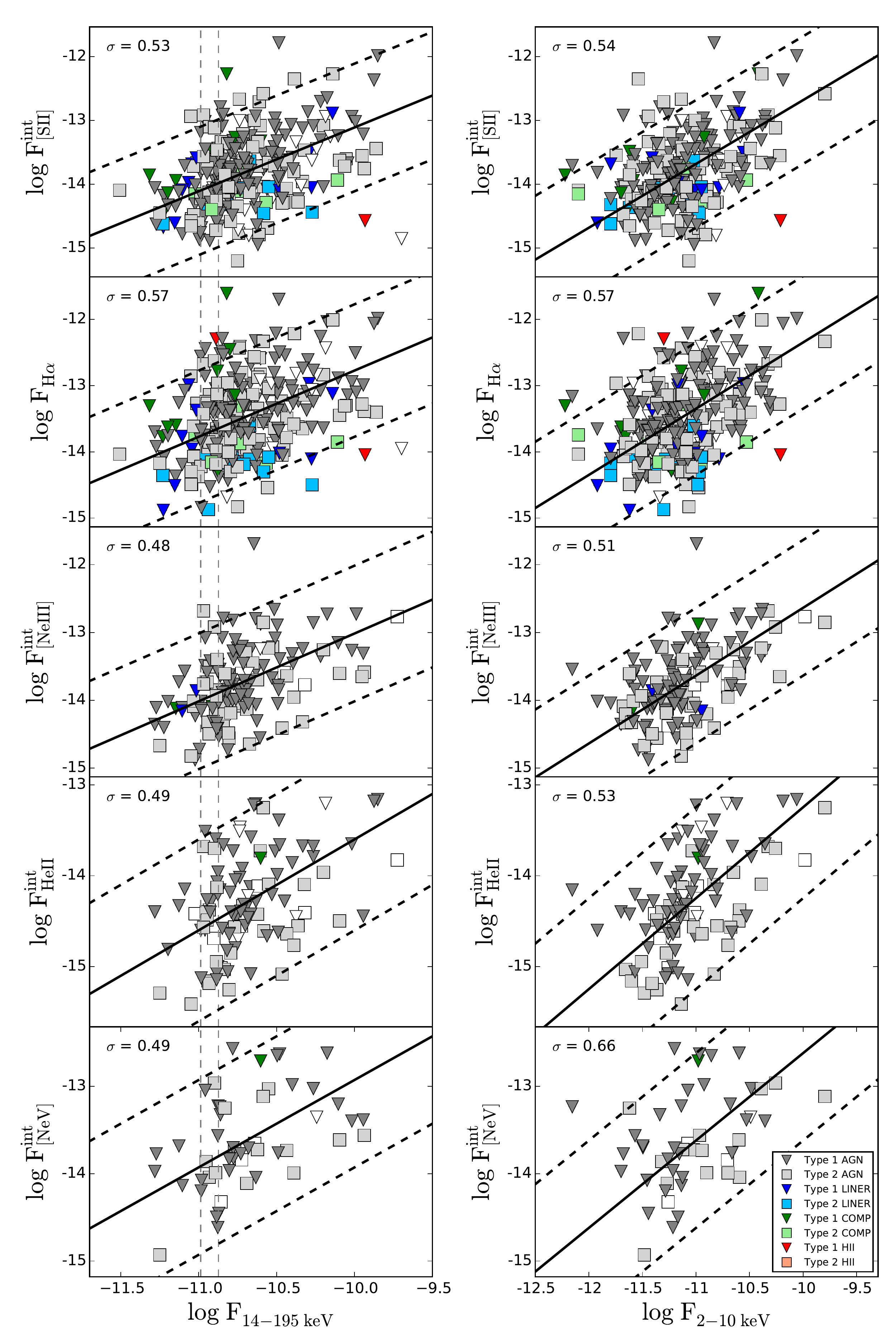}
\caption{Plots of the relation between five optical narrow emission lines (\SII, \ha, \HeIIop, \NeIIIa\ and \NeVb) with the BAT X-ray flux (\textit{left}) and with the intrinsic 2$-$10 \kev\ flux (\textit{right}). The plots show subsamples where the particular emission line is detected. The flux units are in $erg \cdot cm^{-2}\cdot s^{-1}$. The subplots are vertically ordered by ionization level of the emission line. The black lines are one-to-one fits of the data. The color code represents the \nii\ BPT classification. White symbols do not have any BPT classification because of undetected emission lines.}
\label{optxray}
\end{figure*}

Figure \ref{optxray} illustrates the various emission line fluxes as a function of the two X-ray fluxes.
Our data show that the correlation is slightly stronger with \neiii, \heii\ or \nev\ than with \oiii, \ha\ or \sii, especially for the BAT X-ray band. Their scatter is smaller by about $\sim$0.09 dex and the correlation coefficient increases from $\rm{R_{Pear}\sim 0.36}$ to $\rm{R_{Pear}\sim 0.44}$. However, the \nev\ emission line is only detected in optically selected AGNs with very bright emission lines whereas the \oiii\ line is detected in many faint AGN such as the LINERs and composite objects which has the effect to increase the scatter.  To overcome these possible biases, we need to compare the same samples together.
This is what we show in Table \ref{emissionline}. In this table, the reported standard deviations and Pearson coefficients are calculated based on a sub-sample of objects having both the detection of the considered line and an \oiii\ detection.
In order to quantitatively investigate if the flux correlation of X-rays with \sii, \ha, \neiii, \heii\ or \nev\ differ significantly from the one with \oiii, we run a $z$-test based on the two Pearson correlation coefficients (Fisher r-to-z transformation). The $p$-value is also reported in Table \ref{emissionline}.  The result of this statistical analysis shows that we have $p$-value $>$ 0.5 for all emission lines. The small differences between the linear relations are therefore not statistically significant.\\

\subsection[]{Relationships between optical emission lines}
\begin{figure*}
\centering
\includegraphics[width=0.89\textwidth]{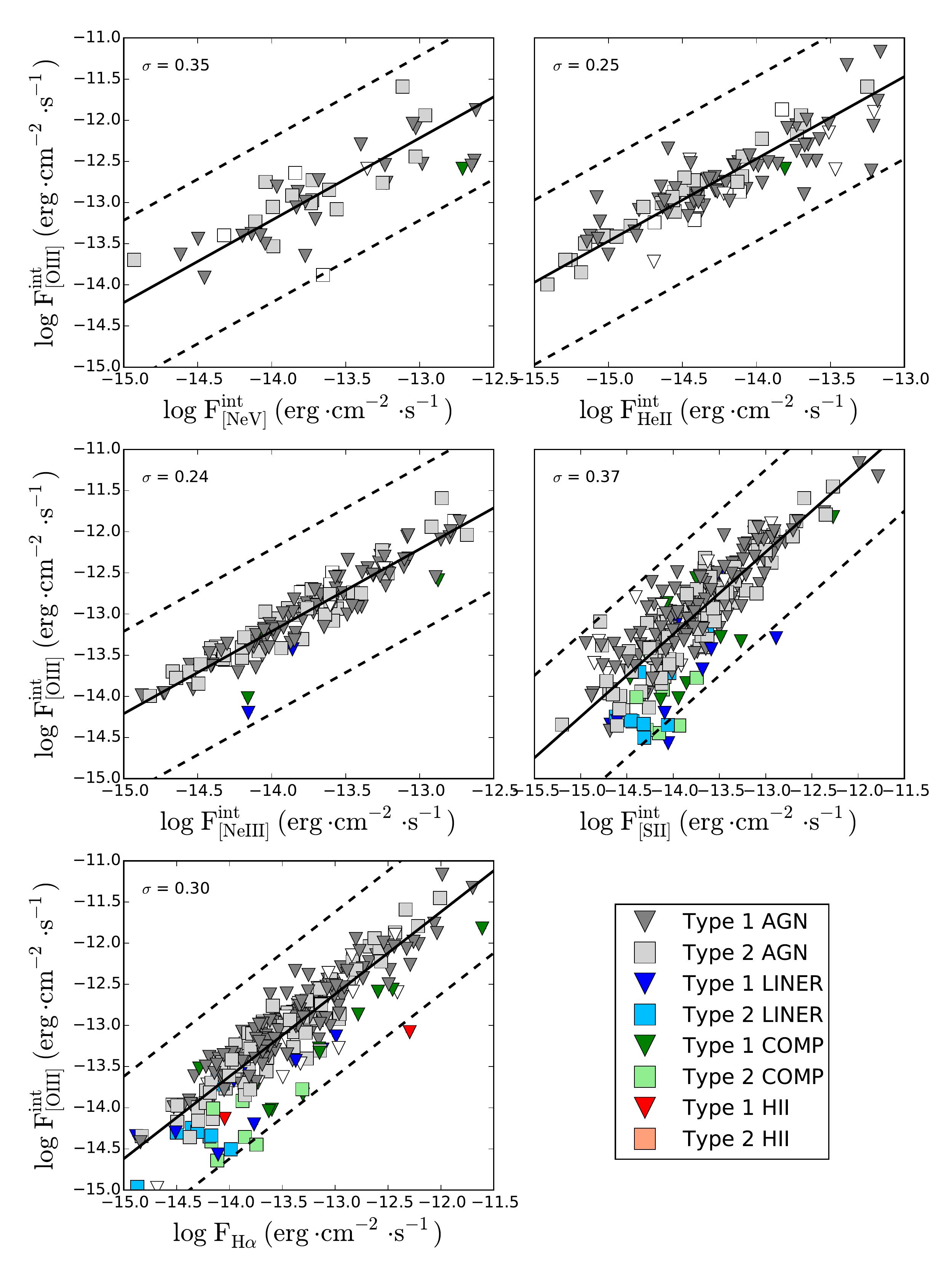}
\caption{Plots of the relation between optical narrow emission lines and the \OIII\ line in flux space. The plots show subsamples where the particular emission line is detected. The lines are all strongly correlated with \oiii. The black lines are one-to-one fits of the data. The color code represents the \nii\ BPT classification. White symbols do not have any BPT classification because of undetected emission lines.}
\label{optopt}
\end{figure*}

Figure \ref{optopt} shows the relation between \oiii\ and other optical emission lines. We see from the scatter ($\sim$0.3 dex) that these emission lines are tightly correlated together. The mean flux ratios, standard deviations and Pearson coefficients are summarized in Table \ref{optical_comp}. One can especially notice the \oiii« -- »\neiii\ relation that exhibits a scatter of $\sigma = 0.24$ dex.
The two transitions originate from ions with comparable ionization energies (35.1 and 41 eV, respectively), and from gas with comparable critical densities \citep[$\mathrm{5 \times 10^5\ cm^{-3}}$ and $\mathrm{7 \times 10^6\ cm^{-3}}$, respectively;][]{electron} which may explain the small scatter.\\


\section[]{Summary and Discussion}
We have presented a study of emission line properties of the largest hard X-ray selected sample of AGN in the local Universe. Our conclusions are:
\begin{enumerate}\itemsep10pt
\item We find very weak correlations with a large scatter (0.62-0.63 dex) between the two X-ray fluxes (2$-$10 \kev\ and 14$-$195 \kev\ band) and the observed and dust corrected \oiii.  The ratios between the X-ray fluxes and the \oiii\ corrected flux are $F_{14-195}=(267\pm 22)\times F_{\oiii}^{int}$ and $F_{2-10}=(102\pm 8)\times F_{\oiii}^{int}$.


\item The correlation of emission lines with high ionization levels (\neiii, \heii, \nev) with the X-ray flux is not significantly better than  for \oiii.  This is also true for the lower ionization line \sii\ and the \ha\ recombination line as compared to \oiii.  While objects which have a \nev, \heii\ or \neiii\ detection have a smaller scatter ($\sigma^{\nev}_{\oiii}=0.49$, $\sigma^{\neiii}_{\oiii}=0.47$, $\sigma^{\heii}_{\oiii}=0.45$),  this is only because they have brighter emission lines that exclude the population of  LINERS or X-ray detected AGN classified as HII/comp.    In general, all lines show a statistically similar level of scatter with the X-rays independent of ionization energy.


\item All the optical emission lines are strongly correlated with \oiii. These correlations ($\sigma \approx 0.3$) are significantly tighter than what we find between \oiii\ and the X-rays.

\item The \oiii/X-ray flux ratio as a function of the column density (\NH) is consistent with being constant. However, we see a strong increase in the scatter above \NH$>10^{22}$ \nhunit\ ($\sigma_{<22}=0.47$, $\sigma_{>22}=0.72$). Our results show that the scatter slightly decreases towards larger physical slit sizes in kpc while the ratio $L_{\oiii}^{int}/L_{X}$ is constant for the AGN in our sample ($z>0.01$).



\end{enumerate}




We find that our study agrees well with past studies of X-ray selected AGN, but shows a weaker correlation than \oiii\ selected samples which exclude X-ray selected AGN with weak emission lines (e.g., LINERs and optically elusive AGN classified as HII/comp). The correlation we find  between the \oiii\ corrected luminosity and BAT X-ray luminosity agrees with the 9-month sample of BAT AGN \cite[e.g., $\rm{R_{Pear}^2 = 0.41}$ vs. $\rm{R_{Pear}^2 \approx }$ 0.35-0.40;][]{lwinter}. Our correlation result disagrees significantly with the 9-month sample of \cite{melendez} which compared 2$-$10 \kev\ X-ray fluxes from the {\em ASCA} database which are not calculated using detailed X-ray modeling.  \cite{panessa_oiii} calculated the correlation between the \oiii\ corrected flux and the 2$-$10 \kev\ flux from a sample of 47 nearby Seyfert galaxies from the Palomar optical spectroscopic survey \citep{Ho_catalog}. They obtained a strong Spearman correlation coefficient of $r_S = 0.78$ that is also not in accordance with this work ($r_S = 0.39$) most likely because their sample was optically selected and did not include AGN that were X-ray weak. Our results show a slightly higher scatter than the \oiii« -- »X-ray relation in \cite{heck05} ($\sigma=0.51$; this work:  $\sigma=0.62$) whose data come from the \textit{RXTE} slew survey (3-20 \kev). \\


We find no significant evidence of a weakening or strengthening of line emission at high AGN luminosities.  This is because of the very weak correlation and large scatter which make a linear model inappropriate (e.g. Hogg et al. 2010).  The improvement in the scatter of a linear fit is very small ($\approx$ 0.01 dex) over a simple one-to-one relation between \oiii\ and X-ray luminosity.  Using the OLS bisector method, the best slope between the intrinsic \oiii\ and the 14$-$195 \kev\ luminosity is $1.29 \pm 0.05$ and the intrinsic \oiii\ and 2$-$10 \kev\ luminosity is $1.23\pm 0.05$.  This is in accordance with previous BAT studies \citep[e.g., $1.16\pm 0.20$;][]{lwinter}.  Alternatively, the EM method including censured data finds a slope of $0.83\pm 0.06$ which is consistent with past studies using the same line fitting method as found in \citealt{melendez} ($0.8\pm 0.1$) or \citealt{panessa_oiii} ($0.82\pm 0.04$).  In summary, our best fit slope is consistent with past studies, though the large intrinsic scatter of the relation yields very different results depending on the line fitting method used.   \\

Our analysis shows that the relation between optical emission lines and hard X-rays exist, but is very weak (at best $\sigma \sim$0.5 dex), no matter which optical emission line (\sii, \ha, \oiii, \neiii, \heii, \nev) or which X-ray flux (2$-$10 \kev\ intrinsic or 14$-$195 \kev).  Additionally, calibration, obscuration, and physical slit size play only a small role in the scatter.     This scatter is  much larger than the scatter of X-ray compared to SED fits of unobscured BAT AGN using the UV \citep[e.g., 14$-$195 keV scatter of 0.3 dex,][]{vasudevan}.  While our analysis focused on the hard X-rays ($>$2 keV), the soft X-rays ($<$2 keV) have been shown to be dominated by emission lines in  several Seyfert AGN whose extended spatial profiles show morphological correlations with the \oiii\ (e.g. Bianchi 2006, Wang et al. 2011, Koss et al. 2015).  Thus, the soft X-rays coming from the emission line regions may shower a tighter correlation that is worth investigating in future studies. \\  


These results therefore strongly suggest that the majority of the intrinsic scatter has physical origins related to the NLR gas.  There are several mechanisms such as covering factor of the narrow-line region gas \citep[e.g.,][]{netzer93,scattering_fraction},  density dependence of \oiii\ \citep{crenshaw05}, differing ionization parameter \citep{baskin2005}, contamination of the narrow emission lines from star formation \citep[e.g.,][]{wild11}, errors in the Balmer decrement because of different distribution of Balmer lines and forbidden lines, and finally SED shape changes with luminosity \citep{netzer2006}.  Another possibility is AGN variability, where the X-ray emission varies on very short time scales (sub-parsec scale) and represents the instantaneous accretion rate while the NLR which is orders of magnitude larger than the corona (100 parsec to kiloparsec scale), varies on much longer time scales and traces the average accretion rate \citep[e.g.,][]{variability, flickering}. Future studies using high signal-to-noise Integral Field Unit (IFU) observations with wide wavelength coverage of large samples of AGN such as the Multi Unit Spectroscopic Explorer \citep[MUSE;][]{muse} are critical to map and study the importance of these different parameters. Finally, the \textit{Nuclear Spectroscopic Telescope Array} \citep[\textit{NuSTAR};][]{nustar} hard X-ray telescope is currently observing many BAT AGN as part of the legacy survey (Balokovi\'{c} et al., in prep.) with a 100 times increase in sensitivity which improves estimates of X-ray luminosity compared to BAT.

\begin{table*}
\centering
\begin{minipage}{140mm}
\caption{\oiii« -- »X-ray luminosity relations for different sub-samples}
\label{tablelum}
\begin{tabular}{|c|c|l|c|c|c|c|c|c|}  \hline \hline
Y & X & Sample & N & a & b & $\sigma$ & $<$X$>$& $<$Y$>$\\
(1) & (2) & (3) & (4) & (5) & (6) & (7) & (8) & (9)\\ \hline


 
 &  & all & 327 & 1.29$\pm$0.05 & -15$\pm$2 & 0.61 &43.9&41.4 \\
 & & Type 1-1.9 & 215 & 1.17$\pm$0.06 & -9.6$\pm$2.6 & 0.53& 44.0 & 41.6 \\
 &  & Type 2.0 & 112 & 1.40$\pm$0.10 & -20$\pm$4 & 0.69 & 43.7 &41.1 \\
 $\log\ L_{\oiii}^{int}$ & $\log\ L_{14-195}$  & & & & & & & \\
  &  &  Seyferts & 241 & 1.16$\pm$0.05 & -9.3$\pm$2.4 & 0.53 & 43.9 & 41.5 \\
 &  &  LINERs & 17 & 1.8$\pm$0.3 & -39$\pm$13 & 0.70 & 43.6 &40.5 \\
 &  &  HII/comp/ambiguous & 17 & 1.2$\pm$0.2 & -11$\pm$8 & 0.74& 43.7 & 40.9 \\ \hline

 &  & all & 321 & 1.23$\pm$0.05 & -12$\pm$2 & 0.61&43.5&41.4 \\
& & Type 1-1.9 & 209 & 1.10$\pm$0.06 & -6.5$\pm$2.5 & 0.56& 43.6&41.6\\
 &  & Type 2.0 & 112 & 1.30$\pm$0.10 & -15$\pm$4 & 0.69&43.3&41.1 \\
$\log\ L_{\oiii}^{int}$ & $\log\ L_{2-10}^{int}$  & & & & & & & \\
 &  &  Seyferts & 238 & 1.13$\pm$0.05 & -7.5$\pm$2.3 & 0.54 & 43.4 & 41.5 \\
 &  &  LINERs & 17 & 1.8$\pm$0.3 & -37$\pm$12 & 0.62 & 43.1 & 40.5 \\
 &  & HII/comp/ambiguous & 16 & 1.0$\pm$0.2 & -4$\pm$8 & 0.62& 43.2 &40.8 \\ \hline
 
\end{tabular}
\medskip
\\Note. « --- »  (1) variable Y ; (2) variable X ; (3) AGN type : type 1-1.9 are all the objects with broad \ha\ detection, type 2 are all the objects without broad detection, LINERs are selected by the \sii\ and \oi\ diagrams, HII/comp/ambiguous and Seyferts are selected by the \nii\ diagram (except for previously selected LINERs); (4) size of sample ; (5) slope with its 1-$\sigma$ uncertainty ; (6) regression intercept with its 1-$\sigma$ uncertainty. The equation of the linear regression is Y = aX + b ; (7) standard deviation ; (8) mean value of X ; (9) mean value of Y.
\end{minipage}
\end{table*}

\begin{table*}
\centering
\begin{minipage}{140mm}
\caption{Relations of various optical emission lines intrinsic fluxes with the 14$-$195 \kev\ emission and the comparison with the \oiii« -- »14$-$195 \kev\ relation.}
\label{emissionline}
\begin{tabular}{l|c|c|c|c|c|c|c} \hline \hline
Line& $\chi$ [eV]& N & $\sigma_{\oiii}$\footnote{scatter of the \oiii« -- »14195 \kev\ relation only for objects with \oiii\ and the other emission line (1) detection to avoid selection biases.} & $\sigma_{Line}$ & $\rm{R_{Pear,\  \oiii}}$\footnote{Pearson correlation coefficient for the same sub-sample used for $\sigma_{\oiii}$} & $\rm{R_{Pear,\  Line}}$ & $p$-value \\
(1)&(2)&(3)&(4)&(5)&(6) &(7)&(8) \\ \hline
\OI & . & 206  & 0.60 dex & 0.53 dex & 0.41 & 0.46 & 0.54\\ \hline
\SII & 10.4 & 275  & 0.61 dex & 0.53 dex & 0.37 & 0.38 & 0.89\\ \hline
\ha\ & 13.6 & 296 & 0.62 dex& 0.57 dex & 0.36 & 0.34 & 0.78 \\ \hline
\NII & 14.5 & 277  & 0.62 dex & 0.56 dex & 0.37 & 0.36 & 0.89 \\ \hline
\NeIII & 41.0 & 144  & 0.47 dex&0.48 dex& 0.45 & 0.44 & 0.92 \\ \hline
\HeIIop & 54.4 & 106 & 0.45 dex&0.49 dex& 0.49 & 0.42 & 0.53 \\ \hline
\NeV & 97.1 & 51 & 0.49 dex& 0.49 dex& 0.42 & 0.45 & 0.86 \\ \hline
\end{tabular}
\\Note. « --- » (1) optical emission line ; (2) ionization level ; (3) size of the common sample in which both the Line and \oiii\ are detected ; (4) standard deviation and (5) Pearson R coefficient of the $\log\ F_{\oiii}^{int}$« -- »$\log\ F_{14-195}$ relation ; (6) standard deviation and (7) Pearson R coefficient of the $\log\ F_{Line}^{int}$« -- »$\log\ F_{14-195}$ relation ; (8) $p$-value of the null hypothesis that the two correlation coefficients obtained from independent parent samples are equal.
\end{minipage}
\end{table*}

\begin{table*}
\centering
\begin{minipage}{140mm}
\caption{Correlation between \OIII\ and other optical emission line intrinsic fluxes.}
\label{optical_comp}
\begin{tabular}{|l|c|c|c|c|c|}  \hline \hline
Line& $\chi$ [eV]& N & $< r >$ & $\sigma$& $\rm{R_{Pear}}$ \\
(1)&(2)&(3)&(4)&(5)&(6) \\ \hline
\OI & . & 206 & 0.050 & 0.36 dex & 0.84 \\ \hline
\SII & 10.4 & 275 & 0.19 & 0.37 dex & 0.83 \\ \hline
\ha & 13.6 & 296 & 0.438& 0.30 dex  & 0.90  \\ \hline
\NII & 14.5 & 277 & 0.30 & 0.40 dex & 0.81 \\ \hline
\NeIII & 41.0 & 138 & 0.17 & 0.24 dex & 0.89  \\ \hline
\HeIIop & 54.4 & 106 & 0.031 & 0.25 dex & 0.88 \\ \hline
\NeV & 97.1 & 43 & 0.18 & 0.35 dex & 0.79  \\ \hline
\end{tabular}
\\Note. « --- » (1) optical emission line ; (2) ionization level ; (3) size of the sample ; (4) mean flux ratio ; (5) standard deviation ; (6) Pearson R coefficient.
\end{minipage}
\end{table*}

\section*{Acknowledgments}
M.\,K. acknowledges support from the Swiss National Science Foundation (SNSF) through the Ambizione fellowship grant PZ00P2\textunderscore154799/1.  M.\,K. and K.\,S. acknowledge support from SNSF grant PP00P2 138979/1.  Support for the work of E.\,T. was provided by the Center of Excellence in Astrophysics and Associated Technologies (PFB 06), by the FONDECYT regular grant 1120061 and by the CONICYT Anillo project ACT1101.  The work of D.\,S. was carried out at Jet Propulsion Laboratory, California Institute of Technology, under a contract with NASA.  M.\,B. acknowledges support from NASA Headquarters under the NASA Earth and Space Science Fellowship Program, grant NNX14AQ07H.  C.\,R. acknowledges financial support from the CONICYT-Chile EMBIGGEN Anillo (grant ACT1101).  Based on observations obtained at the Gemini Observatory, which is operated by the  Association of Universities for Research in Astronomy, Inc., under a cooperative agreement 
with the NSF on behalf of the Gemini partnership: the National Science Foundation (United States), the National Research Council (Canada), CONICYT (Chile), the Australian Research Council (Australia), Minist\'{e}rio da Ci\^{e}ncia, Tecnologia e Inova\c{c}$\widetilde{\text{a}}$o 
(Brazil) and Ministerio de Ciencia, Tecnolog\'{i}a e Innovaci\'{o}n Productiva (Argentina).  Data in this paper were acquired through the Gemini Science Archive and processed using the Gemini IRAF package and Gemini python.  Data from Gemini programs GN-2009B-Q-114, GN-2010A-Q-35, GN-2011A-Q-81, GN-2011B-Q-96, GN-2012A-Q-28, GN-2012B-Q-25, GS-2010A-Q-54, and GS-2011B-Q80 were used in this publication.  We acknowledge the work that the \swift\ BAT team has done to make this work possible. The Kitt Peak National Observatory observations were obtained using MD-TAC time as part of the thesis of M.K. and L. W. at the University of Maryland. Kitt Peak National Observatory, National Optical Astronomy Observatory, is operated by the Association of Universities for Research in Astronomy (AURA), Inc., under cooperative agreement with the National Science Foundation.  Funding for SDSS-III has been provided by the Alfred P. Sloan Foundation, the Participating Institutions, the National Science Foundation, and the U.S. Department of Energy Office of Science.  This research made use of the NASA/IPAC Infrared Science Archive, which is operated by the Jet Propulsion Laboratory, California Institute of Technology, under contract with the National Aeronautics and Space Administration. Funding for the SDSS and SDSS-II has been provided by the Alfred P. Sloan Foundation, the Participating Institutions, the National Science Foundation, the U.S. Department of Energy, the National Aeronautics and Space Administration, the Japanese Monbukagakusho, the Max Planck Society, and the Higher Education Funding Council for England. The SDSS Web Site is http://www.sdss.org/. The SDSS is managed by the Astrophysical Research Consortium for the Participating Institutions. The Participating Institutions are the American Museum of Natural History, Astrophysical Institute Potsdam, University of Basel, University of Cambridge, Case Western Reserve University, University of Chicago, Drexel University, Fermilab, the Institute for Advanced Study, the Japan Participation Group, Johns Hopkins University, the Joint Institute for Nuclear Astrophysics, the Kavli Institute for Particle Astrophysics and Cosmology, the Korean Scientist Group, the Chinese Academy of Sciences (LAMOST), Los Alamos National Laboratory, the Max-Planck-Institute for Astronomy (MPIA), the Max-Planck-Institute for Astrophysics (MPA), New Mexico State University, Ohio State University, University of Pittsburgh, University of Portsmouth, Princeton University, the United States Naval Observatory, and the University of Washington. This research has made use of the NASA/IPAC Extragalactic Database (NED) which is operated by the Jet Propulsion Laboratory, California Institute of Technology, under contract with the National Aeronautics and Space Administration.  This research made use of {\tt Astropy}, a community-developed core Python package for Astronomy (Astropy Collaboration, 2013).  This research made use of {\tt APLpy}, an open-source plotting package for Python hosted at http://aplpy.github.com.  This research has made use of the SIMBAD database, operated at CDS, Strasbourg, France.  This research made use of data and/or software provided by the High Energy Astrophysics Science Archive Research Center (HEASARC), which is a service of the Astrophysics Science Division at NASA/GSFC and the High Energy Astrophysics Division of the Smithsonian Astrophysical Observatory.  This research made use of {\tt pyspeckit}, an open-source spectral analysis and plotting package for Python hosted at http://pyspeckit.bitbucket.org.

\bibliographystyle{mn2e}
\bibliography{ref}

\appendix

\bsp

\label{lastpage}

\end{document}